\documentclass[traditabstract]{aa}
\usepackage{epsfig,graphicx,natbib,txfonts,url,twoopt}
\usepackage[breaklinks=true]{hyperref} 
\usepackage[usenames]{color}

\def\@ #1\par{\par\mbox{}\\ \noindent{\small \tt @ #1}\\[1ex]}  

\bibpunct{(}{)}{;}{a}{}{,}    
\newcommandtwoopt{\citeads}[3][][]{\href{http://adsabs.harvard.edu/abs/#3}%
                                        {\citealp[#1][#2]{#3}}}
\newcommandtwoopt{\citepads}[3][][]{\href{http://adsabs.harvard.edu/abs/#3}%
                                         {\citep[#1][#2]{#3}}}
\newcommandtwoopt{\citetads}[3][][]{\href{http://adsabs.harvard.edu/abs/#3}%
                                         {\citet[#1][#2]{#3}}}
\newcommandtwoopt{\citeyrads}[3][][]{\href{http://adsabs.harvard.edu/abs/#3}%
                                         {\citeyear[#1][#2]{#3}}}

\newcount\longrefs
\def\aap{\ifnum\longrefs=1 {Astron.\ Astrophys.}\else 
                           {A\hbox{\rm \&}A}\fi}
\def\aapr{\ifnum\longrefs=1 {Astron.\ Astrophys.\ Rev.}\else 
                            {A\hbox{\rm \&}AR}\fi}
\def\aaps{\ifnum\longrefs=1 {Astron.\ Astrophys.\ Suppl.}\else 
                            {A\hbox{\rm \&}A Suppl.}\fi}
\def\aj{\ifnum\longrefs=1 {Astron.\ J.}\else 
                          {AJ}\fi} 
\def\ao{\ifnum\longrefs=1 {Applied Optics}\else 
                           {Appl.\ Opt.}\fi} 
\def\aspcs{\ifnum\longrefs=1 {Astron.\ Soc.\ Pacific Conf. Series}\else 
                           {ASP Conf.\ Ser.}\fi} 
\def\apj{\ifnum\longrefs=1 {Astrophys.\ J.}\else 
                           {ApJ}\fi} 
\def\apjl{\ifnum\longrefs=1 {Astrophys.\ J. Lett.}\else 
                            {ApJ}\fi} 
\def\aplett{\ifnum\longrefs=1 {Astrophys.\ J. Lett.}\else 
                            {ApJ}\fi} 
\def\apjs{\ifnum\longrefs=1 {Astrophys.\ J. Suppl.}\else 
                            {ApJS}\fi}
\def\apss{\ifnum\longrefs=1 {Astrophys.\ and Space Science}\else 
                            {Astrophys.\ Space Sci.}\fi}
\def\araa{\ifnum\longrefs=1 {Ann.\ Rev.\ Astron.\ Astrophys.}\else 
                            {ARA\hbox{\rm \&}A}\fi}
\def\azh{\ifnum\longrefs=1 {Astronomicheskii Zhurnal}\else 
                            {Astron.\ Zhur.}\fi}
\def\baas{\ifnum\longrefs=1 {Bull.\ Am.\ Astron.\ Soc.}\else 
                            {BAAS}\fi}
\def\bain{\ifnum\longrefs=1 {Bull.\ Astronom.\ Institutes Netherlands}\else
                            {Bull.\ Astr.\ Inst.\ Neth.}\fi}
\def\gca{\ifnum\longrefs=1 {Geochim.\ Cosmochim.\ Acta}\else 
                           {Geochim.\ Cosmochim.\ Acta}\fi}
\def\grl{\ifnum\longrefs=1 {Geophys.\ Res.\ Lett.}\else 
                           {Geoph.\ Res.\ Lett.}\fi}
\def\iaucirc{\ifnum\longrefs=1 {IAU Circulars}\else 
                          {IAU Circ.}\fi}
\def\ip{\ifnum\longrefs=1 {in press}\else 
                          {in press}\fi}
\def\jgr{\ifnum\longrefs=1 {J.\ Geophys.\ Res.}\else 
                           {J.\ Geophys.\ Res.}\fi}  
\def\jrasc{\ifnum\longrefs=1 {J.\ Royal Astron.\ Soc.\ Canada}\else 
                           {JRAS Can.}\fi}  
\def\memsai{\ifnum\longrefs=1 {Mem.~Soc.~Astron.~Italiana}\else
                              {MemSAI}\fi}
\def\mnras{\ifnum\longrefs=1 {Mon.\ Not.\ Roy.\ Astron.\ Soc.}\else 
                             {MNRAS}\fi} 
\def\nat{\ifnum\longrefs=1 {Nature}\else 
                           {Nat}\fi}
\def\pasj{\ifnum\longrefs=1 {Pub.\ Astron.\ Soc.\ Japan}\else 
                            {PASJ}\fi} 
\def\pasp{\ifnum\longrefs=1 {Pub.\ Astron.\ Soc.\ Pacific}\else 
                            {PASP}\fi} 
\def\physscr{\ifnum\longrefs=1 {Physica Scripta}\else 
                            {Phys.\ Scrip.}\fi} 
\def\planss{\ifnum\longrefs=1 {Planetary \& Space Science}\else 
                            {Plan. \& Space Sci.}\fi} 
\def\procspie{\ifnum\longrefs=1 {Proc.\ SPIE}\else 
                            {Proc.\ SPIE}\fi} 
\def\qjras{\ifnum\longrefs=1 {Quarterly J.\ Royal Astron.\ Soc.}\else 
                            {QJRAS}\fi} 
\def\sa{\ifnum\longrefs=1 {Soviet Astron..}\else 
                               {Sov.\ Astron.}\fi}
\def\skytel{\ifnum\longrefs=1 {Sky \& Telescope}\else 
                            {Sky \& Tel.}\fi} 
\def\solphys{\ifnum\longrefs=1 {Solar Phys.}\else 
                               {Sol.\ Phys.}\fi}
\def\ssr{\ifnum\longrefs=1 {Space Science Rev.}\else 
                               {Space\ Sci.\ Rev.}\fi}
\def\zap{\ifnum\longrefs=1 {Zeitschr.\ f.\ Astrophysik}\else
                               {Z.\ Astrophys.}\fi}

\def\nl{,\ } 

\def\CMAO{Center of Mathematics for Applications, University of Oslo\nl
           P.O. Box 1053, Blindern\nl N-0316 Oslo\nl Norway}

\def\HAO{High Altitude Observatory\nl NCAR\nl PO Box 3000\nl 
         Boulder, CO 80307--3000\nl USA}

\def\ITA{Institute of Theoretical Astrophysics\nl
         University of Oslo\nl
         P.O. Box 1029, Blindern\nl N--0315 Oslo\nl Norway}

\def\SIU{Sterrekundig Instituut\nl Utrecht University\nl Postbus 80\,000\nl
         NL--3508 TA Utrecht\nl The Netherlands}

\long\def\startignore #1\stopignore{}   

\def\rmit#1{{\it #1}}              
\def\etal{\rmit{et al.}}           
           
\def\ie{\rmit{i.e.,}}              
\def\eg{\rmit{e.g.,}}              
\def\cf{cf.}                       

\def\specchar#1{\uppercase{#1}}    

\def\BaII{\mbox{Ba\,\specchar{ii}}}

\def\CaII{\mbox{Ca\,\specchar{ii}}} 
\def\CaIII{\mbox{Ca\,\specchar{iii}}}

\def\FeI{\mbox{Fe\,\specchar{i}}}

\def\HI{\mbox{H\,\specchar{i}}} 
 
\def\MgI{\mbox{Mg\,\specchar{i}}}

\def\NaI{\mbox{Na\,\specchar{i}}}

\def\SiII{\mbox{Si\,\specchar{ii}}}

\def\Halpha{\mbox{H\hspace{0.1ex}$\alpha$}} 

\def\Hbeta{\mbox{H\hspace{0.2ex}$\beta$}}

\def\NaDone{\mbox{Na\,\specchar{i}\,\,D$_1$}}
\def\NaDtwo{\mbox{Na\,\specchar{i}\,\,D$_2$}}

\def\Done{\mbox{D$_1$}}

\def\Mgbtwo{\mbox{Mg\,\specchar{i}\,b$_2$}}

\def\HK{\mbox{H\,\&\,K}}

\def\KtwoV{\mbox{K$_{2V}$}}
\def\KtwoR{\mbox{K$_{2R}$}}

\def\HtwoV{\mbox{H$_{2V}$}}

\def\rmb{{\rm b}}              

\def\kms{\hbox{km$\;$s$^{-1}$}}

\def\is{\!=\!}                             

\def\={\hbox{$\!=\!$}}                     

\def\rmit#1{#1}               
\def\specchar#1{{\sc #1}}     
\def\kms{km/s}    

\long\def\rev#1{{{\bf #1}}}   
\long\def\rev#1{#1}           

\def\figspath{../figs}
\def\figspath{.}

\begin{document}

\title{Quiet-Sun imaging asymmetries in 
  Na\,I D$_1$ compared with other strong Fraunhofer lines}
\subtitle{}
\titlerunning{Imaging asymmetries in strong Fraunhofer lines}

\author{R.J. Rutten\inst{1,2}
        \and
        J. Leenaarts\inst{1}
        \and
        L.H.M. Rouppe van der Voort\inst{2}
        \and
        A.G. de Wijn\inst{3}
        \and
        M. Carlsson\inst{2,4}
        \and
        V. Hansteen\inst{2,4}}

\authorrunning{R.J. Rutten \etal}

\institute{\SIU \and \ITA \and \HAO \and \CMAO}

\date{Received; accepted}


\abstract{ Imaging spectroscopy of the solar atmosphere using the
  \NaI\ \Done\ line yields marked asymmetry between the blue and red
  line wings: sampling a quiet-Sun area in the blue wing displays
  reversed granulation, whereas sampling in the red wing displays
  normal granulation.  The \Mgbtwo\ line of comparable strength does
  not show this asymmetry, nor does the stronger \CaII\,8542\,\AA\
  line.  We demonstrate the phenomenon with near-simultaneous spectral
  images in \NaI\ \Done, \Mgbtwo, and \CaII\,8542\,\AA\ from the
  Swedish 1-m Solar Telescope.  We then explain it with line-formation
  insights from classical 1D modeling and with a 3D
  magnetohydrodynamical simulation combined with NLTE spectral line
  synthesis that permits detailed comparison with the observations in
  a common format.  The cause of the imaging asymmetry is the
  combination of correlations between intensity and Dopplershift
  modulation in granular overshoot and the sensitivity to these of the
  steep profile flanks of the \NaI\ \Done\ line.  The \Mgbtwo\ line
  has similar core formation but much wider wings due to larger
  opacity buildup and damping in the photosphere.  Both lines obtain
  marked core asymmetry from photospheric shocks in or near strong
  magnetic concentrations, less from higher-up internetwork shocks
  that produce similar asymmetry in the spatially averaged
  \CaII\,8542\,\AA\ profile.}

\keywords{Sun: photosphere 
       -- Sun: chromosphere 
       -- Sun: surface magnetism
       -- Sun: faculae, plages}

\maketitle


\section{Introduction}                              \label{sec:introduction}
Three technological advances currently improve optical studies of the
solar atmosphere.  First, adaptive optics and numerical postprocessing
deliver angular resolution in groundbased imaging well beyond the
atmospheric Fried limit (rarely larger than 10~cm equivalent aperture
even at the best sites).  Second, multi-line imaging spectroscopy
leaped forward with fast-tuning narrow-band Fabry-P\'erot
interferometers at the major solar telescopes.  Third, numerical MHD
simulations have evolved into productive analysis tools of small-scale
near-surface magnetoconvection and are becoming reliable emulators of
fine structure higher up in the solar atmosphere.

In this paper we combine these advances in a comparative study of the
formation of three of the strongest lines accessible to the current
generation of Fabry-P\'erot instruments, which all observe in the red
part of the spectrum.  The three lines are the b$_2$ line of \MgI\ at
$\lambda \is 5172.698$\,\AA\ (henceforth ``Mg line''), the D$_1$ line
of \NaI\ at $\lambda \is 5895.940$\,\AA\ (henceforth ``Na line''), and
the infrared-triplet line of \CaII\ at $\lambda \is 8542.144$\,\AA\
(henceforth ``Ca line'').  The wavelengths are for standard air and
taken from \citetads{1966sst..book.....M}. 

These three lines are often taken to be diagnostics of the solar
chromosphere, together with \Halpha\ which gave the chromosphere its
name through its off-limb spectral dominance (\citeads{Lockyer1868};
see \citeads{Rutten2010b}). 
On the solar disk \Halpha\ displays dense \rev{canopies of fibrils
  that cover active regions,} network, and even quiet-Sun internetwork
cell interiors except in the very quietest regions.  The detailed
comparison between \Halpha\ and the Ca line by
\citetads{2009A&A...503..577C} 
shows that the latter line displays these chromospheric fibrils
similarly near \rev{quiet-Sun} network but with smaller extent,
whereas in \rev{quiet} internetwork cell centers it samples the domain of
generally cool but shockwave-ridden gas below the fibrilar canopies
that was termed ``clapotisphere'' by
\citetads{1995ESASP.376a.151R}. 
We use this name here also, reserving ``chromosphere'' for the fibril
canopies seen in \Halpha.  The Na core is mostly formed within the
wave-dominated clapotispheric domain, as shown by
\citetads{2010ApJ...709.1362L}. 
We show below that the same holds for the Mg core.

The present study is a companion to the observational \Halpha\,--\,Ca
line comparison of
\citetads{2009A&A...503..577C} 
by now comparing the Na and Mg lines to the Ca line.  It is a direct
extension of the observation--simulation comparisons by
\citetads{2009ApJ...694L.128L} 
for the Ca line and by \citetads{2010ApJ...709.1362L} 
for the Na line, by now comparing these lines with another and with
the Mg line.  However, in this study the emphasis does not lie on the
chromospheric or clapotispheric response of the cores of these lines
but rather on their inner-wing formation in layers only 100-200~km
kilometer above the surface, where the phenomenon of ``reversed
granulation'' dominates the observed scene (\eg\
\citeads{1962ApJ...135..474L}; 
\citeads{1964ApNr....9...33E}; 
\citeads{1972SoPh...27..299E}; 
\citeads{1973SoPh...33...33C}; 
\citeads{1976ApJ...203..533A}; 
\citeads{1984ssdp.conf..181N}; 
Suemoto \etal\
\citeyrads{1987SoPh..112...59S}, 
\citeyrads{1990SoPh..127...11S}; 
\citeads{2004A&A...416..333R}; 
\citeads{2005A&A...431..687L}; 
\citeads{2006A&A...450..365J}; 
\citeads{2007A&A...461.1163C}). 
The topic of this paper is not its physical nature, but how how this
domain appears in the three lines.

\def\figlabel{fig:august}    
\begin{figure*}       
  \centering
  \includegraphics[width=18cm]{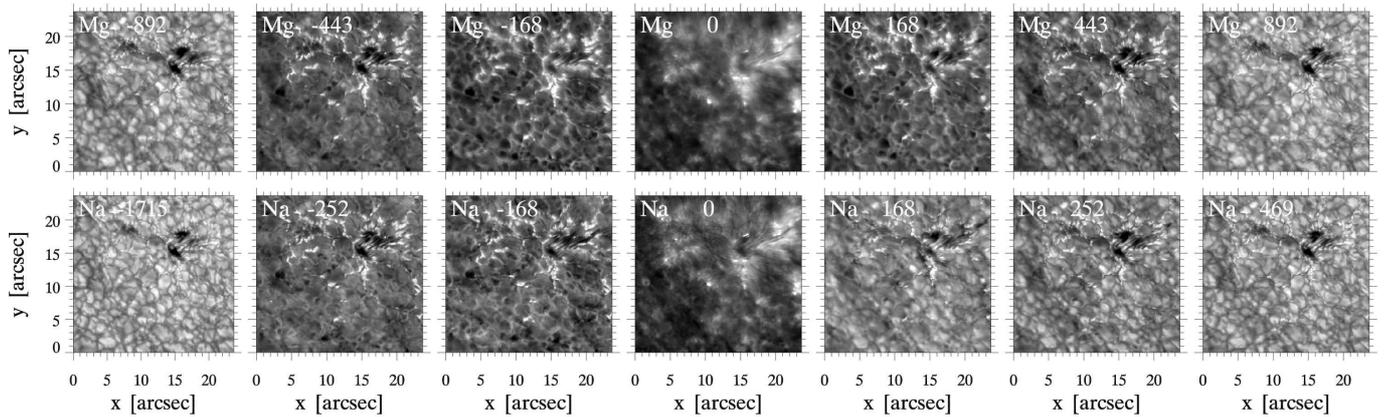} 
  \caption[]{\label{\figlabel} %
    Profile-sampling image comparisons between the Mg line ({\em upper
      row\/}) and the Na line ({\em lower row\/}).  These images are
    subfield cutouts from a dual-line SST/CRISP profile scan taken on
    August 23, 2010.  They sample a small active region with pores in
    the upper part and a quiet area in the lower part of the subfield.
    Each image is byte-scaled independently and is labeled with its
    wavelength separation $\Delta \lambda$ from line center in m\AA.
    Both sequences sample the line profile from the outer blue wing
    through the core to the outer red wing.  The $\Delta \lambda$
    values in the first two columns were selected for scene similarity
    between the two lines.  The next two columns were also selected
    for scene similarity but have equal $\Delta \lambda$ values for
    the two lines.  The final three columns mirror the $\Delta
    \lambda$ values of the first three.  All images show rich detail;
    we invite the reader to magnify them with a pdf viewer.  This
    paper explains the imaging asymmetry between the two lines in the
    red-wing panels where the Mg line shows reversed granulation as in
    its blue wing, whereas the Na line instead shows normal granulation 
    as in the last panel.}
\end{figure*}

This study started with the data shown in Figure~\ref{fig:august} from
the CRISP imaging spectrometer at the Swedish 1-m Solar Telescope
(SST, La Palma).  They sample the Mg and Na lines from a small active
region.  We first inspected the line-center images in the fourth
column.  They display bright fine-scale fibrilar structure near some
pores, next to dark internetwork elsewhere.  Thus, the Mg and Na cores
do display some chromospheric response, \ie\ can be opaque in fibrils.
These line-center images are remarkably similar for the two lines.

We then compared the images taken in the blue wings of the two lines
and quickly found that for every scene observed in the one line, a
nearly identical scene can be found in the other.  One only has to
search further out into the much stronger Mg wing.  We concluded that
these wings sample the same solar layers similarly, and that there is
no point in observing them both.

However, when we then turned to the red wings we were surprised by the
striking difference illustrated in Figure~\ref{fig:august}: in the red
Na wing the reversed granulation drops out.  The Mg line shows
reversed granulation in its red wing that is similar to that in its
blue wing, but in the Na red wing the line-center scene of
chromospheric activity is quickly replaced by granulation, without an
intermediate stage exhibiting reversed granulation.  Along the red Na
wing one seems to suddenly drop from chromospheric and clapotispheric
sampling at line center to photospheric granulation sampling further
out.

These inspections were made at the SST while T.D.~Tarbell (LMSAL, Palo
Alto) and P.~S\"utterlin (ISP, Stockholm) were present.  The first
commented that he had noted this reversed/non-reversed granulation
asymmetry between the Na line wings in Hinode/SOT images and had
wondered about its cause.  The second suggested granular
brightness-Dopplershift correlation as in Figure~6 of
\citetads{2001A&A...378..251S}. 

Their suggestions were followed up in an observing campaign at the SST
during October 2010, targeting truly quiet Sun to better view reversed
granulation and adding the Ca line because it must sample the same
scenes somewhere along its extended wings.  We analyze the best
resulting data here, comparing them to modeling predictions from a
three-dimensional MHD simulation snapshot following Leenaarts \etal\
(\citeyrads{2009ApJ...694L.128L}, 
\citeyrads{2010ApJ...709.1362L}). 

Below, we describe the observations in Section~\ref{sec:observations},
the simulation and the spectral synthesis in
Section~\ref{sec:simulations}, and the results in
Section~\ref{sec:results}. Since the latter consist of
profile-sampling image cubes for the three lines for both observations
and simulations, we use a common format permitting direct comparison
to inspect the observed and simulated cubes in a parallel descriptive
characterization of what they show.  We then start the interpretation
in Section~\ref{sec:interpretation} by reviewing what one expects from
mean-profile inspection and from NLTE line synthesis with a standard
one-dimensional hydrostatic solar-atmosphere model. We use these
insights to interpret the results and explain most observed features.
We add Hinode images as indicator of center-limb behavior, and end the
paper with discussion in Section~\ref{sec:discussion} and conclusions
in Section~\ref{sec:conclusion}.

\section{Observations}                            \label{sec:observations}

\subsection{SST/CRISP observations}
The CRisp Imaging SpectroPolarimeter (CRISP,
\citeads{2008ApJ...689L..69S}) 
at the Swedish \mbox{1-m} Solar Telescope (SST,
\citeads{2003SPIE.4853..341S}) 
contains a dual Fabry-P{\'e}rot tunable filter system.  It is designed
for diffraction-limited narrow-band imaging with fast passband tuning
sampling multiple spectral lines at rapid cadence in the wavelength
range 5100--8600~\AA.

The observing campaign, which was educationally oriented involving
Oslo students on-site, suffered from bad observing conditions.
However, on the last day, October 30, 2010, there was a 40-min period
of moderate-quality seeing permitting observation of a region close to
disk center \rev{(Stonyhurst heliographic coordinates (N4, W0) degrees)}
that was selected for its absence of any rosettes or other
chromospheric activity in \Halpha.

The observing sequence sampled four spectral lines: the Mg, Na and Ca
lines discussed here and also \FeI~6302\,\AA\ at $\Delta \lambda \is
-0.048$\,\AA\ from line center to obtain magnetograms.  The Mg line
was sampled at 41 wavelengths between $\Delta \lambda = \pm
0.892$\,\AA\ from line center with a passband with $\mbox{FWHM} \is
32$\,m\AA, the Na line at 41 wavelengths between $\Delta \lambda = \pm
1.715$\,\AA\ with passband 57\,m\AA, the Ca line at 47 wavelengths
spanning $\Delta \lambda = \pm 2.717$\,\AA\ with passband 111\,m\AA.
The sampling had regular spacing across each core but was sparser in
the outer wings.  The scan durations were about 10~s per line.  At
every wavelength step 8 exposures were acquired.  Since the cameras
operated at 35~frames/s speed and the prefilter change time ranged
from 0.27~s to 0.52~s, the total duration for one complete scan
through all 4 lines was 29.7~s.  We selected the best scan for this
analysis.  It was taken from 09:09:03 to 09:09:33~UT.

The spatial resolution of these observations was improved beyond the
seeing Fried limit by using the SST adaptive optics system described
by \citetads{2003SPIE.4853..370S} 
and by subsequent postprocessing with the Multi-Object Multi-Frame
Blind Deconvolution (MOMFBD) technique of
\citetads{2005SoPh..228..191V}. 
More details on such image reduction are given in, \eg\
\citetads{2008A&A...489..429V} 
and \citetads{2009ApJ...705..272R}. 
The field of view after this processing and co-alignment is $58 \times
56$~arcsec$^2$, with 0.0592~arcsec/px or 42.6~km/px scale.
However, the figures below show only a smaller area which equals the
horizontal simulation extent.

\subsection{Conversion to brightness temperature} \label{sec:brighttemp}
The brightness temperature $T_\rmb$ is the formal temperature that
yields $B_\lambda(T_\rmb) = I_\lambda^{\rm obs}$, \ie\ that reproduces
the observed intensity when entered into the Planck function. It
equals the gas temperature $T_{\rm gas}$ at the height where
$\tau_\lambda=1$ along a radial line of sight when LTE and the
Eddington-Barbier approximation are valid.  LTE does not hold for the
cores of our lines, but we nevertheless convert intensities into
brightness temperatures to enable comparisons that are not
affected by the differences in Planck function sensitivity to temperature
at different wavelengths.

The inverse Planck operation requires rescaling of the observed
intensities into absolute units.  We did this by comparison to the
solar disk-center spectrum atlas obtained by Brault and Testerman with
the Fourier Transform Spectrometer (FTS) at Kitt Peak, calibrated by
\citetads{1984SoPh...90..205N} 
and spread by \citet{Neckel1999}. 
We smeared the atlas profiles of the three lines with the
corresponding CRISP spectral transmission functions, \ie\ the product
of the measured prefilter transmission and the theoretical
transmission of each of the two etalons at the line wavelengths.  The
resulting functions have a prominent symmetrical sidelobe pattern,
with the first pair reaching 0.001~relative amplitude at $\Delta
\lambda = \pm 2.2$\,\AA\ for the Na line.  Convolving the atlas
profiles with these makes the line cores rise substantially.  For each
line, we rescaled the data units of the spatially-averaged observed
profile to the absolute intensities of the smeared atlas profile by
determining their ratio for the wavelength with the highest outer-wing
intensity in the observed profile.  The observed and the smeared atlas
cores show satisfactory agreement (last column of
Figure~\ref{fig:sampleprofiles}), confirming that the transmission
functions give an adequate description of the spectral smearing.

\subsection{Hinode observations}
The Hinode satellite
(\citeads{2007SoPh..243....3K}) 
was programmed to observe at several \rev{center-to-limb} positions
along the central meridian on December 4, 2010 between 14:14:33 and
15:33:41~UT using the Narrowband Filter Imager (NFI) fed by the 0.5~m
Solar Optical Telescope (SOT, \citeads{2008SoPh..249..167T};
\citeads{2008SoPh..249..197S}).
Successive scans sampling the spectrum in $80~\mathrm{m\AA}$ steps
from $-480$ to $+480~\mathrm{m\AA}$ from line center were made through
the Mg line and the Na line, with bandwidths 74 and 84\,m\AA,
respectively.  One line scan took approximately 22~s.  The prefilter
changes and context images taken with the Broadband Filter Imager
added 10~s delay between successive line scans.  The data were
processed with the standard SolarSoft {\em fg\_prep} pipeline.  We
selected small subfields of the size of the simulation extent from the
pointings at $\mu \is 1.0$, $0.5$, and $0.35$, \rev{where $\mu \!\equiv\!
  \cos \theta$ with $\theta$ the viewing angle,} for display in
Section~\ref{sec:Hinode}.


\def\figlabel{fig:contrasts}    
\begin{figure}       
  \centering
  \includegraphics[width=88mm]{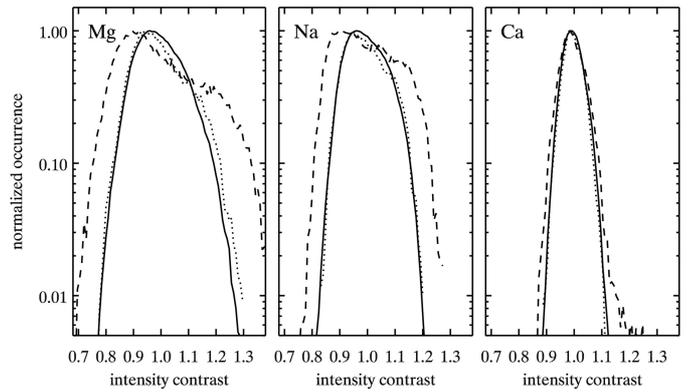} 
  \caption[]{\label{\figlabel} %
    Intensity contrast comparisons for the three lines between
    observations ({\em solid\/}), simulation ({\em dashed\/}), and the
    smeared simulation ({\em dotted\/}).  Each curve shows the
    normalized occurrence distribution of pixels with the given
    contrast (intensity divided by the spatial mean) for the sum of
    the blue and red outermost-wing image pairs.
  }
\end{figure}

\def\figlabel{fig:magnetograms}    
\begin{figure}       
  \centering
  \includegraphics[width=70mm]{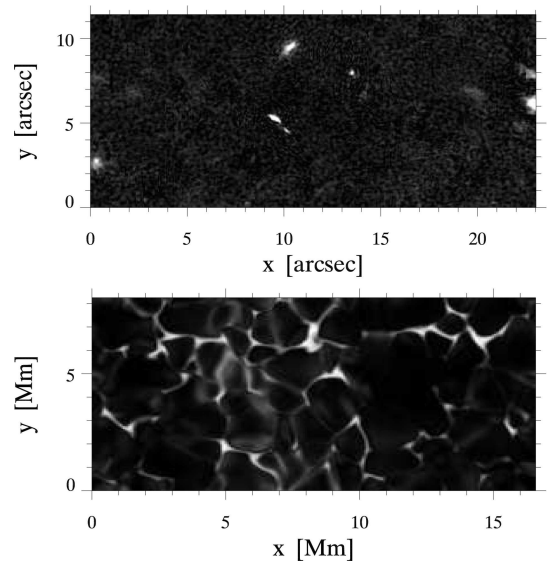} 
  \caption[]{\label{\figlabel} %
    Unsigned magnetic field distributions.  {\em Top\/}: observed
    Stokes-$V$ amplitude at $\Delta \lambda \is -0.048$\,\AA\ in the
    \FeI\,6302\,\AA\ line for the subfield.  Only a few small magnetic
    concentrations have signal above the noise.  The two major ones
    near $x \is 10$ have opposite polarity. %
    {\em Bottom\/}: field amplitude in the simulation at $h \is 0$~km,
    corresponding to spatially-averaged $\tau_5 \is 1$ in the
    continuum at $\lambda \is 5000$\,\AA.  The field is bipolar, as
    shown in Figure~5 of
    \citetads{2010ApJ...709.1362L}. 
    Throughout this paper image scales are in arcsec, simulation
    scales in Mm.  The two areas have equal solar-surface extent.  }
\end{figure}

\def\figlabel{fig:obssamples}  
\begin{figure*}      
  \sidecaption
  \includegraphics[width=12cm]{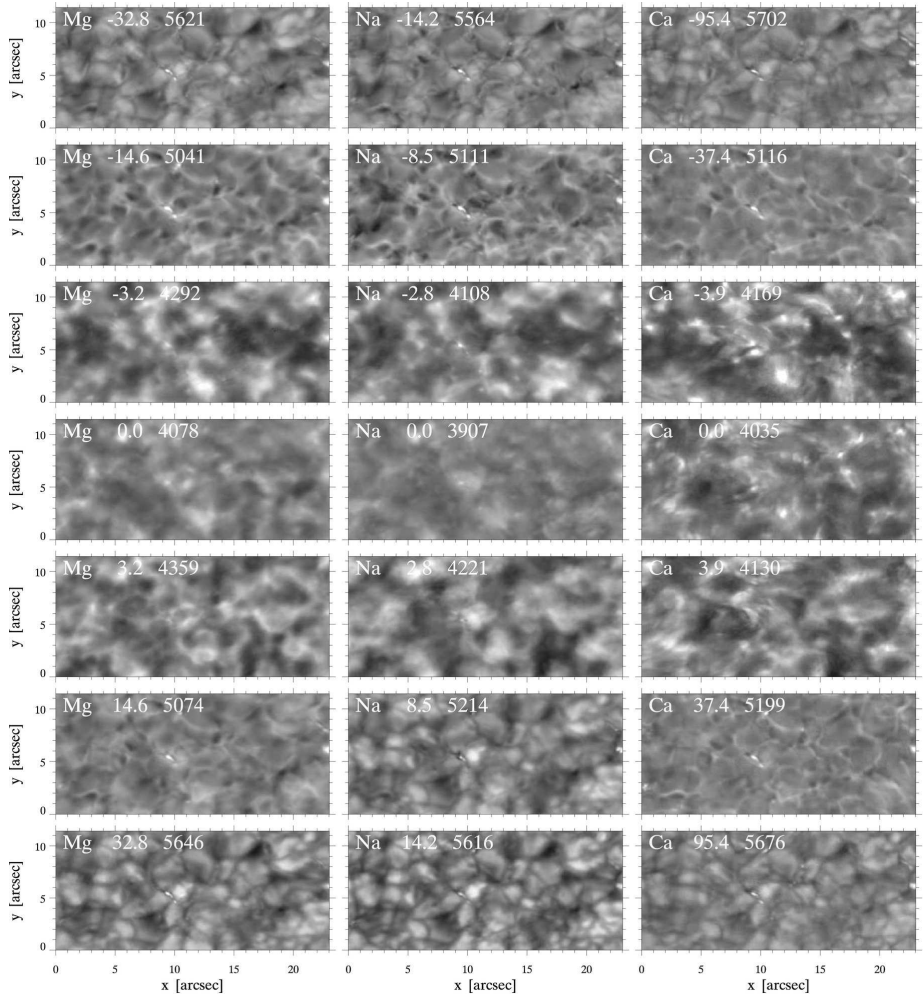}
  \caption[]{\label{\figlabel} %
    Observed brightness-temperature images for selected wavelengths
    sampling the profiles of the Mg line (first column), Na line
    (second column), and Ca line (third column).  The area is a small
    cutout of the center of the full field of view that corresponds in
    size to the horizontal extent of the numerical simulation.  The
    identifier in each panel specifies the line, the CRISP passband
    offset from line center in \kms\ with redshift positive, and the
    mean brightness temperature over the full field of view.  The
    grayscale range of each panel is clipped at $\Delta T_b \is \pm
    400$~K relative to this mean, so that each panel covers the same
    brightness temperature range in magnitude.  The passband offsets
    select equal FALC formation height across the first and last two
    rows (optical depth unity near $h \is 65$ and 150~km,
    respectively) and near-equal wavelength offset $\Delta \lambda$
    from the nominal line center wavelength in Dopplershift units
    across the middle three rows (about $\Delta \lambda \is -3$, $0$,
    and $+3$~\kms, respectively). The selections are symmetrical
    with respect to line center (central row).\\[1ex]}
\end{figure*}

\def\figlabel{fig:simsamples}  
\begin{figure*}      
  \sidecaption
  \includegraphics[width=12cm]{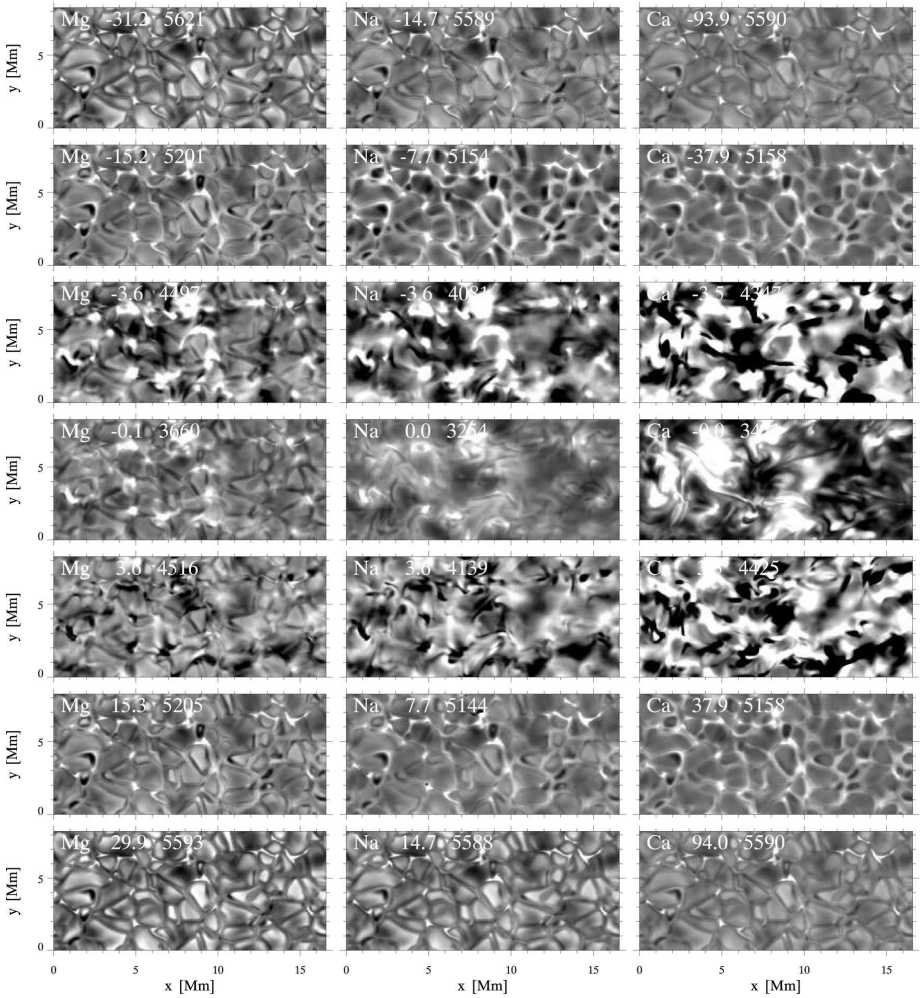}
  \caption[]{\label{\figlabel} %
    Simulated brightness-temperature images for selected monochromatic
    wavelengths sampling the profiles of the Mg line (first column),
    Na line (second column), and Ca line (third column).  The area
    equals the observed subfield in Figure~\ref{fig:obssamples} in
    size.  The format, labels and grayscales correspond to those of
    Figure~\ref{fig:obssamples}, \ie\ each panel is again clipped at
    $\Delta T_b \is \pm 400$~K relative to the spatially averaged
    mean.  The wavelength selection, again specified in Dopplershift
    units, corresponds to the selection in
    Figure~\ref{fig:obssamples}, \ie\ near-equal FALC formation height
    of about 65~km and 150~km along the outer rows and near-equal
    wavelength offset $\Delta \lambda$ from the nominal line center in
    Dopplershift units across the middle three rows, of which the
    center one samples nominal line center. \\[1ex] }
\end{figure*}

\def\figlabel{fig:simcrispsstsamples}  
\begin{figure*}      
  \sidecaption
  \includegraphics[width=12cm]{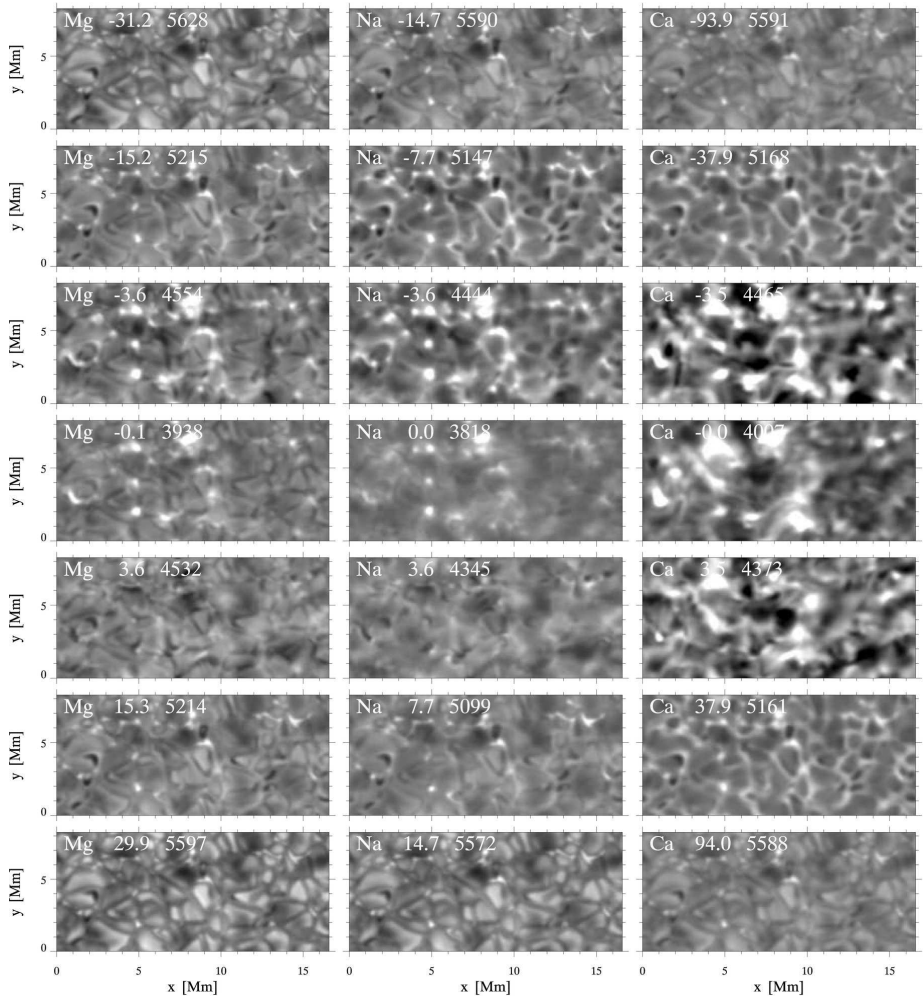}
  \caption[]{\label{\figlabel} %
    The same simulated brightness-temperature images as in
    Figure~\ref{fig:simsamples} but after spectral smearing with the
    CRISP transmission profiles described in
    Section~\ref{sec:brighttemp} and spatial smearing with the point
    spread function estimates obtained in Section~\ref{sec:smearing}.
    \\[1ex]}
\end{figure*}

\def\figlabel{fig:doppsamples}  
\begin{figure*}      
  \sidecaption
  \includegraphics[width=12cm]{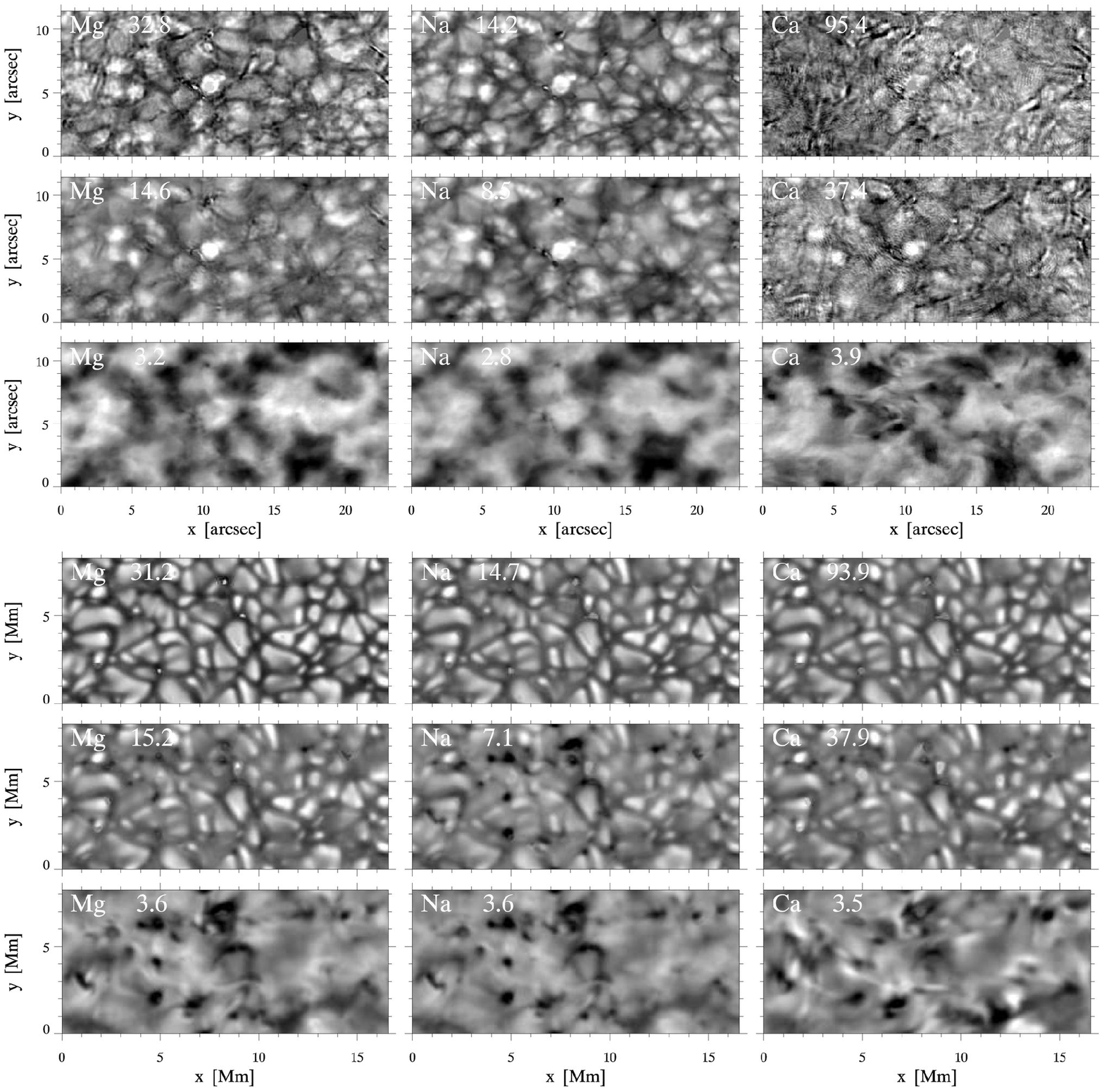}
  \caption[]{\label{\figlabel} %
    {\em Upper assembly\/}: observed brightness-temperature
    Dopplergrams for the three lines in the format of
    Figure~\ref{fig:obssamples}.  Each panel displays the normalized
    Dopplergram subtraction $D \equiv (T_\rmb^{\rm red} - T_\rmb^{\rm
      blue})/ (T_\rmb^{\rm red} + T_\rmb^{\rm blue})$, with bright
    implying blueshift (updraft) for spectral absorption features and
    redshift (downdraft) for spectral emission features.  The
    grayscale is defined so that $D \is 0$ is represented by the same
    gray in all panels while the ranges are adapted per panel for
    largest contrast.  The separations from line center $\Delta
    \lambda$, specified in \kms\ in each panel, are the same as in
    Figure~\ref{fig:obssamples}. %
    {\em Lower assembly\/}: corresponding results from the smeared
    simulation.  The $\Delta \lambda$ values are the same as in
    Figures~\ref{fig:simsamples} and
    \ref{fig:simcrispsstsamples}. \\[1ex]}
\end{figure*}

\section{Simulations}                              \label{sec:simulations}

\subsection{Simulation snapshot}
We employ the same snapshot from a 3D radiation-MHD simulation
performed with the Oslo Stagger Code
(\citeads{2007ASPC..368..107H}) 
that was used in
\citetads{2009ApJ...694L.128L} 
for Ca line synthesis and in
\citetads{2010ApJ...709.1362L} 
for Na line synthesis.  This simulation snapshot contains significantly more
magnetic field than the observations (Figure~\ref{fig:magnetograms}),
but we prefer to re-use the snapshot and spectral line synthesis
nevertheless because otherwise we would have to re-do and re-display
much analysis and diagnosis presented in Leenaarts \etal\
(\citeyrads{2009ApJ...694L.128L}, 
\citeyrads{2010ApJ...709.1362L}) 
that we can now simply refer too.

The $(x,y,z)$ snapshot grid measures $256 \times 128 \times 160$
sampling points, corresponding to a physical size of $16.6 \times 8.3
\times 15.5$~Mm$^3$ with the height ranging from the upper convection
zone to the corona. The grid spacing is 65~km/px in the horizontal $x$
and $y$ directions;
in the vertical $z$ direction it varies from 32~km at the bottom to
440~km at the top.  The snapshot contains bipolar magnetism with a
mean field strength of 150~gauss in the photosphere.  The vertical
field magnitude at the bottom of the photosphere is shown in
Figure~\ref{fig:magnetograms}.  Figure~5 of
\citetads{2010ApJ...709.1362L} 
displays various physical quantities, including field strength, at
different heights in the simulation.

\subsection{Radiative transfer}
For the Ca line and the Na line we also re-use the line synthesis
described in Leenaarts \etal\
(\citeyrads{2009ApJ...694L.128L}, 
\citeyrads{2010ApJ...709.1362L}), 
respectively, to which we refer for further detail.  These radiative
transfer computations were performed in 3D geometry using the code
Multi3D of \citetads{1997MsT..........2B} 
and \citetads{2009ASPC..415...87L} 
which is based on the 1D code MULTI of
\citetads{1986UppOR..33.....C}. 
For this paper, simpler profile synthesis of the Mg line was added by
using MULTI and assuming each vertical column of the simulation
snapshot to describe the stratifications of a plane-parallel 1D
atmosphere.

Because 3D radiative transfer computation is very computationally
demanding, the model atom for Na described in
\citetads{2010ApJ...709.1362L} 
contains only the levels governing the line, the continuum, and an
intermediate level with large collision bound-free coupling to ensure
the ``photon suction'' population flow that tends to populate the
\NaI\ ground state (\citeads{1992A&A...265..237B}). 
The model atom for Ca was a simple 5-level plus continuum one which is
adequate to describe the formation of \CaII\ \HK\ and the \CaII\
infrared triplet including our Ca line (\eg\
\citeads{1989A&A...213..360U}). 
The model atom for Mg used here was a 13-level plus continuum model
that was also used by \citeads{2010MmSAI..81..777F} 
and is based on the 65-level plus continuum model atom of
\citetads{1992A&A...253..567C}. 
The level reduction used the method of
\citetads{2008ApJ...682.1376B}. 
Photoionization was evaluated with line-blocking accounted for by
opacity sampling in 9000 wavelength points as described by
\citeads{2005A&A...442..643C}. 

\subsection{Spectral and spatial smearing}  \label{sec:smearing}
In order to mimic the observations we first smeared the simulation
intensities in the spectral domain with the CRISP transmission
functions discussed in Section~\ref{sec:brighttemp}.  Their effect is
large in the line cores and small in the outer wings
(Figure~\ref{fig:sampleprofiles}).

Subsequently, we smeared the simulation spatially with an estimated
SST point spread function (PSF).  Its determination is less
straightforward.  Even the shape of the typical PSF remains in
debate.  \citetads{1971A&A....14...15L} 
adopted the sum of two Gaussians, as did
\citetads{1975A&A....45..167D}, 
but \citetads{1984ssdp.conf..174N} 
advocated the sum of two Lorentzians instead.
\citetads{1987A&A...180..223C} 
tested both formalisms and concluded that the choice strongly affects
granulation contrast values that result from restoration by
deconvolution.
\citetads{2003ApJ...597L.173S} 
used the sum of the SST's theoretical Airy function and a Lorentzian
to describe the atmospheric degradation, while the comparable paper by
\citetads{2004A&A...427..335S} 
smeared the same simulation with the sum of the DOT's theoretical
Airy function and a two-parameter wing function following Equation~6
of \citetads{1987A&A...180..223C}, 
which is not a Lorentzian but describes the straylight across a
knife-edge such as the lunar limb when the PSF is Lorentzian.
\citetads{2008A&A...487..399W} 
and \citetads{2009A&A...503..225W} 
defined PSFs for seeing-free imaging with Hinode/SOT as the
convolution of the theoretical Airy function and a Voigt function, 
best-fitting the Voigt parameters with Mercury transit and eclipse
data.

In our case both seeing and scattering cause angular smearing, we have
no eclipse or transit data to measure their effects, and we must cope
with the wavelength variation between our lines.  We therefore do not
aim to obtain a reliable PSF for data deconvolution, but simply apply
somewhat realistic smearing to the simulation as illustration of
its likely effect.  We chose to use a single Voigt profile,
representing the convolution of the Gaussian approximation to the Airy
function $A(r) \approx 1/(\sqrt{2\pi}\,w)\,\exp(-r^2/2w^2)$ with
extended Lorentzian tails $L(r) =
\gamma/(4\,\pi^2)/[(\gamma/4\pi)^2+r^2]$ describing scattering, and
assume circular symmetry with radius $r$ as variable.  In principle,
the Airy width is $w=0.42\,\lambda/D$, but since the seeing during
this observation was only mediocre (in terms of the superb seeing that
the La Palma site offers at its best), the MOMFBD restoration probably
did not reach the full-aperture resolution.  We therefore reduce the
aperture $D$ entering the Airy component to an ``effective'' one with
a fractional factor $f$ and scale the resulting $w \propto
\lambda^{-1/5}$ as given by atmospheric turbulence theory (see
\citeads{1981PrOpt..19..281R}). 
The Voigt parameter $a=\gamma/(4\pi\sqrt{2}\,w)$ is scaled with
$\lambda^{s}$. We thus have three fit parameters: $f$, $a$, and $s$.

As fit constraints we used the resulting contrast distribution and the
visual appearance of the granulation at the extremes of our common
wavelength ranges, \ie\ the deepest-formed wavelengths that permit
observation-simulation comparison.  These differ per line.  We use
these extremes because the simulation is probably most realistic at
the granulation level (indeed, most recent solar PSF estimates aim to
show that granulation simulations are more reliable than
PSF-challenged observations).  We sum these outer wing images as a
first-order compensation for Dopplershifts. (It would be better to use
parallel nearby-continuum observations but the CRISP prefilters do not
permit taking such.)  We found that the combination of $f \is 0.5$ and
$a \is 0.2$ at the wavelength of the Mg line with $s \is -2$ produces
both the acceptable contrast fits shown in Figure~\ref{fig:contrasts}
and similar appearance of the observed and smeared granulation (not
shown, but compare the outer rows of Figures~\ref{fig:obssamples} and
\ref{fig:simcrispsstsamples}).  The value $f \is 0.5$ seems realistic
for non-superb seeing; $a \is 0.2$ adds extended wings below the 1\%
level, and $s \is -2$ describes scattering between the Rayleigh and
Mie limits.  The differences in distribution width between panels in
Figure~\ref{fig:contrasts} primarily reflect different Planck function
sensitivities to temperature.  The dip and enhanced righthand tail,
most clearly present for the Mg line, are due to the appreciably
larger amount of magnetism in the simulation.

\section{Results}                                    \label{sec:results}

In this section we present displays from the profile-sampling image
cubes for the three lines from the observations and from the
simulation, likewise and in a common format.  These figures serve to
supply a comparative empirical inventory of what quiet-Sun fine
structures the three lines display, in parallel for observation and
simulation, while postponing all interpretation to
Section~\ref{sec:interpretation}.  We again invite the reader to
magnify the figures per pdf viewer for better appreciation of their
detail.

\subsection{Magnetograms}  \label{sec:magnetograms}
Figure~\ref{fig:magnetograms} compares the magnetic field in the
subfield of the observations and in the simulation.  The observed
magnetogram indicates apparent longitudinal flux density, the
simulation panel displays intrinsic vertical field strength.  The
observed subfield has only a few very small magnetic concentrations;
there are some larger ones elsewhere in the full field of view.  The
field in the simulation is shown quantitatively in Figure~5 of
\citetads{2010ApJ...709.1362L}, displaying the polarity distribution
in the first panel and the lateral expansion of the magnetic
concentrations with height in subsequent panels.  The larger white
patches in the simulation panel represent kilogauss concentrations.
The few observed white patches are likely to also represent kilogauss
fields. There is much more strong field in the simulation than in the
observation, but it is nevertheless a relatively weak-field
simulation; the field does not upset granular convection as is the
case in strong network or plage.
  
\subsection{Profile-sampling images}  \label{sec:samples}
Figure~\ref{fig:obssamples} shows a selection of the observed
line-sampling images, cut down to a small subfield at the center of
the SST field of view of which the $23.1 \times 11.5$~arcsec$^2$ area
equals the horizontal simulation extent.  All panels have dynamic
range $\Delta T_b \is \pm 400$~K around the mean value over the full
field of view.  The conversion into brightness temperature and the
common greyscale range make all panels intercomparable.  Similarly,
the wavelength separations from line center $\Delta \lambda$ are
specified in Dopplershift units (\kms) to supply
wavelength-independent comparison.

The $\Delta \lambda$ selections are specified in each panel and are
symmetric along the columns around the line-center panels in the
central row.  They have been carefully selected to obtain similar
atmospheric sampling in the three lines.  For the first and second
rows (and reversely the last and next-to-last rows) the $\Delta
\lambda$ values are those that produce equal $\tau_\lambda \is 1$
formation height in the FALC modeling described in
Section~\ref{sec:FALC} below, at $h \is 65$~km and $h \is 150$~km,
respectively.  The first height is the lowest at which we can do such
three-line comparison, being limited by the outermost CRISP tunings in
the extended wings of the Ca line.  This height samples the top of the
normal granulation.  The second height was selected to sample reversed
granulation.  At both heights LTE is a good assumption
(Section~\ref{sec:FALC}); indeed, the mean $T_b$ values specified in
each panel are similar along these rows and between opposite-wing
pairs.  Closer to line center we instead selected similar $\Delta
\lambda$ values along rows, since these images may primarily sample
clapotispheric structures in this very quiet area.  Cloud-like line
formation with specific cloud Doppler velocities should then produce
similar scenes for lines with similar cloud opacities and source
function variations.

The granulation appears closely similar along the first row, but not
so along the bottom row where the Na line displays the largest
granulation contrast, the Ca line the smallest.  The reversed
granulation also appears closely similar in all three lines along the
second row, but not so in the next-to-last row where the Na line
roughly shows the same granulation as in the bottom panel, whereas the
Ca line displays the closest similarity to its blue-wing counterpart
in the second row.  These differences confirm the imaging asymmetries
in Figure~\ref{fig:august}.

The line core scenes in the central three rows are similar for the Mg
and Na lines, but differ for the Ca line.  For all three lines they
differ much on the opposite sides of line center, with opposite
grayscale patterns especially for Mg and Na.  These patterns are
patchy in morphology and appear fuzziest in the Na line.  In the Ca
line small roundish bright features appear, clearest on the blue side
but not on the red side.

The few magnetic concentrations in this very quiet area
(Figure~\ref{fig:magnetograms}) appear as bright points, clearest in
the second row (all three lines) and next-to-last row (Mg and Ca).

Figure~\ref{fig:simsamples} presents the companion display from the
simulation.  The $\Delta \lambda$ selections are the closest available
to those in Figure~\ref{fig:obssamples}.  In all panels the fine
structure has appreciably larger brightness-temperature contrast than
in Figure~\ref{fig:obssamples}, particularly in the central three
line-core rows.  The granulation has largest contrast along the top
and bottom rows in the Mg line, smallest in the Ca line.  The reversed
granulation is most clear for Na in the second row, least clear for Na
in the next-to-last row.  In the latter it appears present as well,
but all the bright patches and lanes in this panel represent strong
magnetic field (compare with Figure~\ref{fig:magnetograms} and the
bottom Na panel).

\def\figlabel{fig:bscatter} 
\begin{figure*}      
  \sidecaption
  \includegraphics[width=12cm]{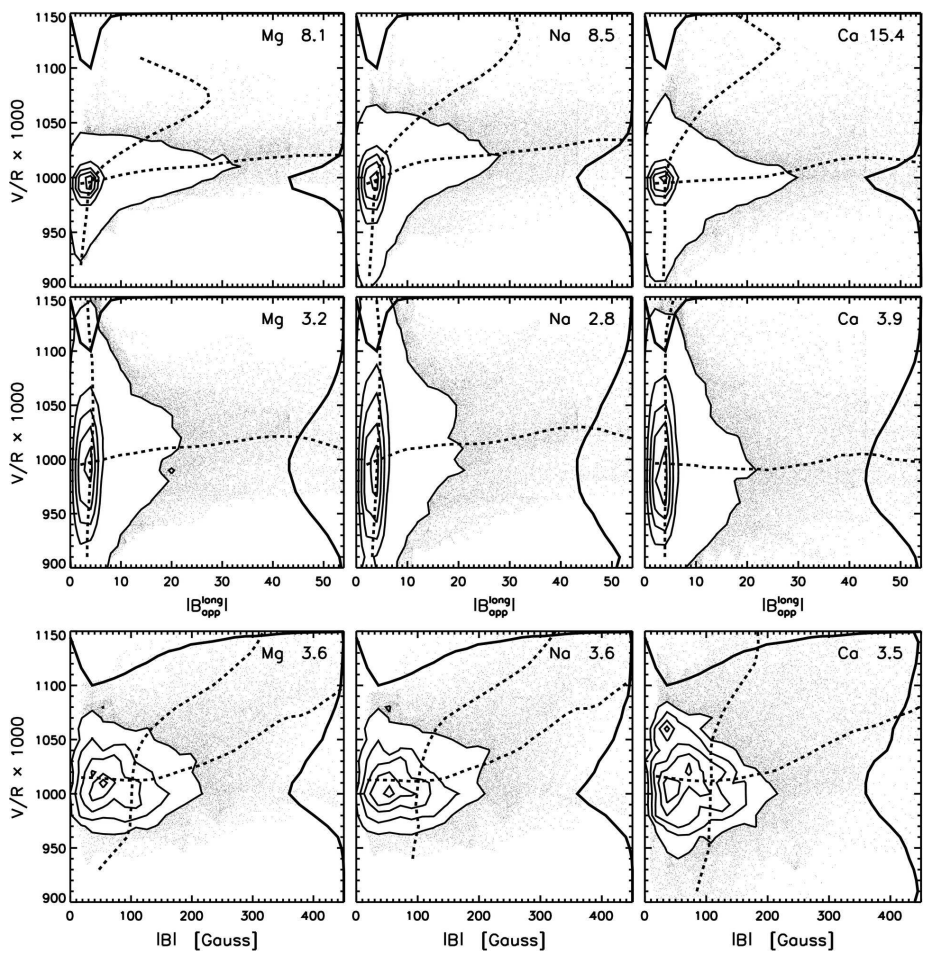}
  \caption[]{\label{\figlabel} %
    Scatter diagrams of the core asymmetry ratio $V/R \equiv
    T_\rmb^{\rm blue} / T_\rmb^{\rm red}$ against field strength.
    {\em Upper two rows\/}: observations, using the full field of view.
    The abscissa is the magnitude of the apparent longitudinal flux
    density measured from the \FeI\,6302\,\AA\ magnetogram, in
    arbitrary units and after 9-pixel spatial smoothing to mimic field
    spreading with height.  {\em Bottom row\/}: smeared simulation.  The
    abscissa is the magnitude of the vector field at height
    $z\is500$~km.  {\em Left\/}: Mg line.  {\em Center\/}: Na
    line. {\em Right\/}: Ca line.  The sampling wavelength in \kms\ is
    specified in each panel.  The $V/R$ values are multiplied by 1000
    for plotting reasons.  Density contours replace individual pixels
    to avoid plot saturation.  The solid curves are the distributions
    over $V/R$ and field strength, respectively.  The dashed curves
    show the first moments per bin per axis.  They line up for high
    correlation and are perpendicular at no correlation. \\[1ex]}
\end{figure*}

In the three line-core rows of Figure~\ref{fig:simsamples} the Mg line
displays dark intergranular lanes in all three rows, most strikingly
at $\Delta \lambda \is +3.6$~\kms.  They are covered by higher-lying
structures in the other two lines.  These are grayish with much fine
structure at Na line center but become black-and-white clipped at the
400~K display limits at Ca line center.  They break up into smaller
patches in the near-center panels, many opposite in black or white at
the two sides.

Figure~\ref{fig:simcrispsstsamples} repeats
Figure~\ref{fig:simsamples} after smearing.  The simulation pixels
(65~km) are actually coarser than the SST pixels (42.6~km on the
observing date), but comparison of Figures~\ref{fig:obssamples} and
\ref{fig:simsamples} suggests that the actual resolution of the
observations is considerably worse.  We therefore apply spectral
smearing with CRISP's transmission profiles and spatial smearing with
the estimated PSFs from Section~\ref{sec:smearing} to illustrate their
effect.  The central three rows of Figure~\ref{fig:simcrispsstsamples}
are most affected, especially with respect to the extended patches in
the Na line. \rev{At line center, these loose the fine structure seen
  in Figure~\ref{fig:simsamples} and become smoother greyish patches,
  appreciably darker than the bright magnetic concentrations and with
  similar appearance as the extended patches in the corresponding
  panel of Figure~\ref{fig:obssamples}.}  Dark intergranular lanes
remain visible in the Mg core.  The Na-line imaging asymmetry of
reversed granulation in the second and next-to-last rows remains, but
without showing dark lanes outside the field-filled ones in the red
wing as observed in Figure~\ref{fig:obssamples}.

\def\figlabel{fig:sampleprofiles}  
\begin{figure*}      
  \centering
  \includegraphics[width=18cm]{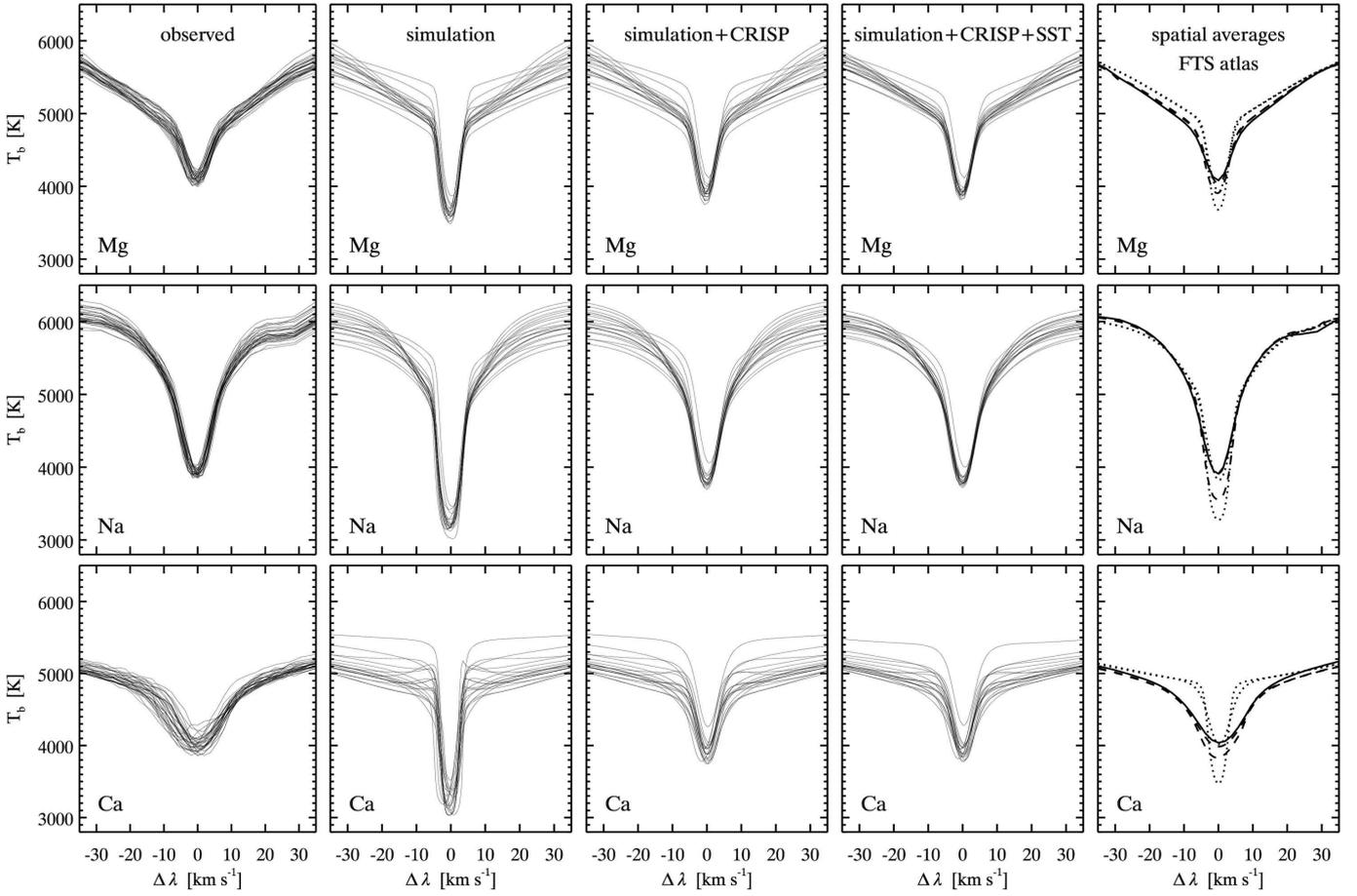}
  \caption[]{\label{\figlabel} %
    Sample profiles of the cores of the Mg line ({\em top row\/}), Na
    line ({\em second row\/}), and Ca line ({\em third row\/}), for
    every 50th pixel in $x$ and $y$ in Figures~\ref{fig:obssamples}
    and \ref{fig:simsamples}, respectively.  {\em First column:\/}
    observed profiles.  {\em Second column:\/} original synthetic
    profiles from the simulation.  {\em Third column:\/} synthetic
    profiles spectrally smeared with the CRISP transmission profile.
    {\em Fourth column:\/} additional smearing with the estimated SST
    point spread function. The smearing was applied to the computed
    intensities, not the brightness temperatures. {\em Final
      column:\/} mean profiles.  The solid curves are the observed
    full-field means, the dotted curves the simulation means without
    smearing and with CRISP and SST smearing (the latter does not
    affect the spatial mean).  The dashed curves are the FTS atlas
    profiles without and with CRISP smearing.  All panels cover the
    same coordinate ranges. \\[1ex]}
\end{figure*}

\subsection{Dopplergrams} \label{sec:Dopplergrams}
The various asymmetries above between the blue and red line wings
obviously suggest a role of Dopplershifts.  Therefore,
Figure~\ref{fig:doppsamples} shows Dopplergrams, in the upper half for
the observations and in the lower half for the smeared simulation.
The sign of the Dopplergram definition $D \equiv (T_\rmb^{\rm red} -
T_\rmb^{\rm blue})/ (T_\rmb^{\rm red} + T_\rmb^{\rm blue})$ is chosen
to make the granular Dopplershift pattern appear as granulation, \ie\
with bright granular updrafts and dark intergranular downdrafts.  The
$D$ ranges are optimized per panel, but $D \is 0$ is the same dark
gray in each panel.  The $\Delta \lambda$ selections correspond to
those in Figures~\ref{fig:obssamples}--\ref{fig:simcrispsstsamples}.
The first rows again sample the top of the normal granulation, the
second rows the reversed granulation, the third rows the inner line
cores.

In the observed Dopplergrams the pattern quality along the first row
shows that the Na line has the best granular Doppler sensitivity, Ca
the worst.  Most structure in the Ca Dopplergram is noise or MOMFBD
artifact that is much exaggerated by the greyscale optimization.  In
the Mg and Na panels some granules have larger positive (bright)
Dopplershift than others.  The reversed granulation sampled in the
second row shows the same updraft pattern as the granulation in the
first row, but the downdraft lane pattern is less clear than the
updraft granule pattern.  

The observed Mg and Na Dopplergrams in the third row have essentially
three shades.  The bright and dark patches (apart from the magnetic
concentrations which appear as small black grains) correspond roughly
to the areas with bright and less bright granules in the upper rows,
respectively.  The intermediate dark-gray background consists of
unsharp granules (more clearly when we flip rapidly through adjacent
Dopplergrams on our screen).  These are not seen in the Ca core
Dopplergram, in which the patches also have different patterning as
well as finer texture.  In this Dopplergram black grains appear also
away from the magnetic concentrations.  The strong updraft in the
granule near the center of the field is visible as a roundish bright
patch in the Mg and Na cores but not in the Ca core.

The simulation Dopplergrams also show the granulation pattern as
dominant in the first row, a much fuzzier granulation pattern with the
lanes washed out in the second row, and larger-scale patch patterns in
the third row.  The fine texture in the latter is similar for the Mg
and Na lines, less so for the Ca line.  The dark gray shade again
corresponds to underlying granules in Mg and Na but not in Ca.  They
are less distinct than in the Mg-core brightness images in
Figure~\ref{fig:simcrispsstsamples}.

When we scan through all observed Dopplergrams, bright updraft patches
appear first at $\Delta \lambda \approx 8$, 6, 15~\kms\ for the Mg,
Na, Ca lines, respectively.  Towards the line centers they become more
opaque.  In the simulations they appear only at $\Delta \lambda
\approx 4$, 5, 5~\kms, respectively.  The simulation Ca Dopplergrams
show a very flat gray, nearly granule-free low-Dopplershift
intermediate scene over $\Delta \lambda \is 11$--6~\kms\ that is not
sampled in the observations.

The more numerous magnetic concentrations in the simulation also stand
out as black blobs in the lower Mg and Na panels.  The dark patches in
the Ca Dopplergram in the last panel have some but less good
correspondence to the strongest fields.  

\subsection{Core asymmetries} \label{sec:asymmetries}
The scatter diagrams in Figure~\ref{fig:bscatter} demonstrate the
field-Dopplershift correlations in more detail.  They plot the core
asymmetry ratio $V/R \equiv T_\rmb^{\rm blue} / T_\rmb^{\rm red}$
against field strength per pixel, both for the observations and for
the simulation.  (We use symbol $V$ for ``violet'' to correspond to
the \CaII\ \HK\ practice of \eg\
\citetads{1983ApJ...272..355C}, 
and to avoid confusion with the $B$ for magnetic field).  The sampling
wavelengths for Mg and Na in the first and bottom rows are those that
show the tightest high-high correlations (line-up of the two moment
curves).  These lie further out in the observed profiles than in the
synthetic profiles.  The observed Mg and Na diagrams in the first row
show significant correlation between large $V/R$ and large field
strength.  The observed Ca line shows similar correlation at double
$\Delta \lambda$ from line center.  The diagrams in the center row
show that the high-high correlations remain similar close to line
center (tilt of the horizontal first-moment curve), but many more
weak-field pixels now contribute larger $V/R$.  For all three
lines the low-field mountain spreads out in $V/R$, with an extended
distribution tail for the Na and Ca lines.  

The scatter diagrams from the simulation (bottom row of
Figure~\ref{fig:bscatter}) have very different field distributions
corresponding to Figure~\ref{fig:magnetograms}.  They show similar
correlations as the observations in the top row, but only close to
line center.  The $V/R$ spreading and distribution tails are more like
the observed ones near line center (center row).

\subsection{Line profiles} \label{sec:profiles}
Figure~\ref{fig:sampleprofiles} concludes our parallel inspection of
observation and simulation results.  It shows random samples of
observed line profiles in the first column, random samples of the
simulated line profiles in the next three columns, and spatially
averaged observed and simulated profiles together with FTS-atlas
profiles in the last column.  The spread between the sample profiles
is much larger for the simulation than for the observation, but
becomes similar with the CRISP and SST smearing.  The simulated cores
rise considerably with the CRISP smearing but do not become as shallow
as the observed ones.  The largest discrepancy is obvious in all
simulation panels, and also in the last column: even after the CRISP
smearing, the simulated line cores are narrower than the observed
ones, not much for the Na line but markedly for the Ca line.  This
difference matches the differences in core-structure onset with
$\Delta \lambda$ and core asymmetry sampling noted above.


\def\figlabel{fig:neckelred}    
\begin{figure}       
  \centering
  \includegraphics[width=70mm]{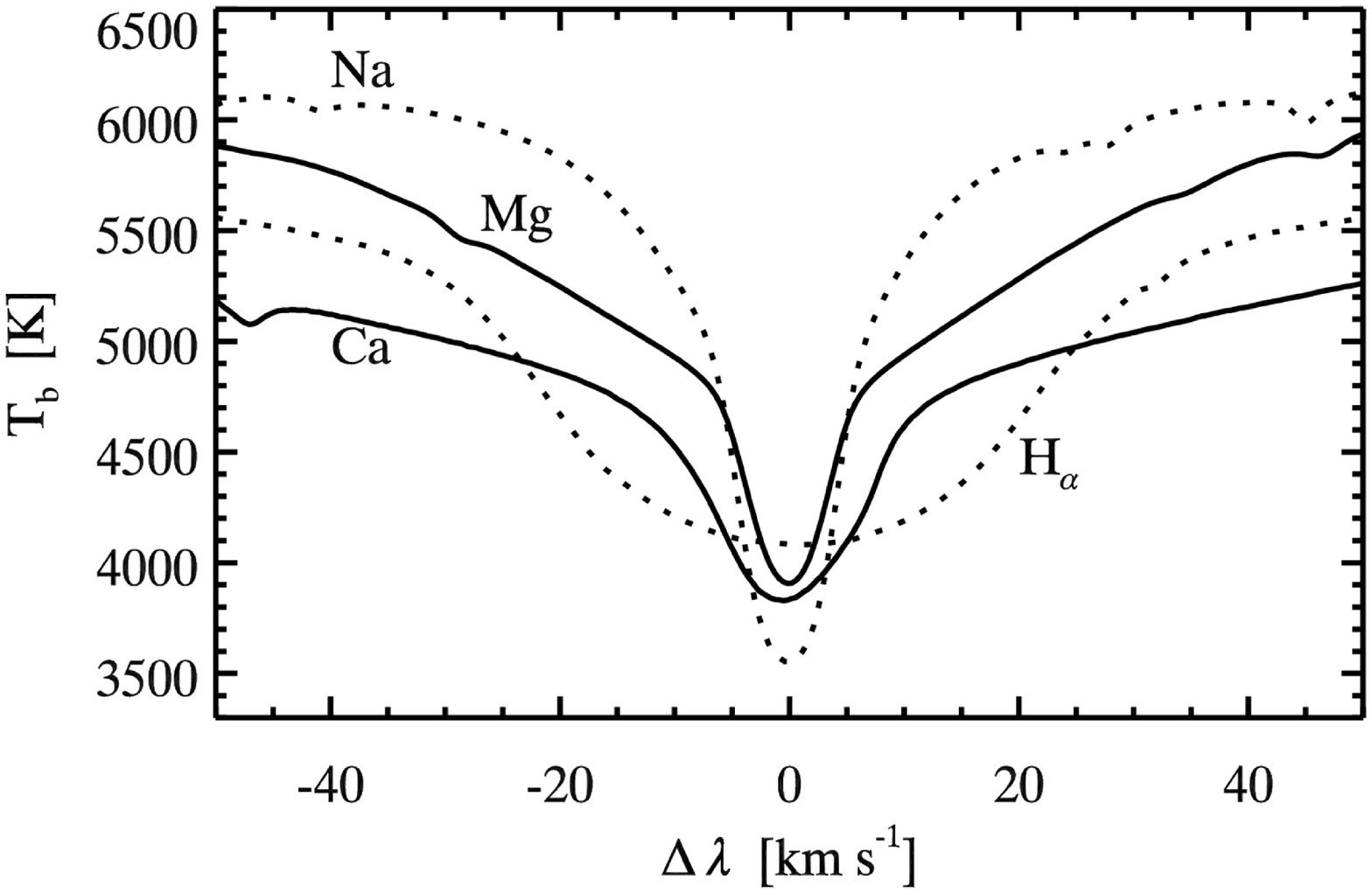} 
  \caption[]{\label{\figlabel} %
    Strong lines in the red part of the solar spectrum.  Disk-center
    profiles of the Na line (\NaI\ 5895.940\,\AA; dotted), the Mg line
    (\MgI\ 5172.698\,\AA), the Ca line (\CaII\ 8542.144\,\AA) and
    \Halpha\ (\HI\ 6562.808\,\AA, dotted) from the FTS atlas are shown
    on common scales.  The atlas calibration by
    \citetads{1984SoPh...90..205N} 
    was used to convert the intensities into formal brightness
    temperatures.  The wavelengths were converted into formal
    Dopplershifts from line center, with redshift positive.  }
\end{figure}

\def\figlabel{fig:neckelcaii}    
\begin{figure}       
  \centering
  \includegraphics[width=70mm]{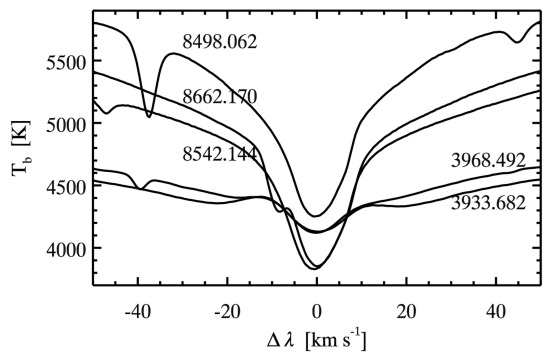} 
  \caption[]{\label{\figlabel} %
    Disk-center profiles of the five strongest \CaII\ lines from the
    FTS atlas.  Format as for Figure~\ref{fig:neckelred}.}
\end{figure}

\section{Interpretation}                        \label{sec:interpretation}

\subsection{Solar atlas profiles}  \label{sec:neckel}
We start with a comparison of the spatially-averaged profiles of the
three lines.  Figure~\ref{fig:neckelred} combines them with \Halpha\
in a common-scale display of their profiles in the FTS atlas.  The Na
line is the deepest, so simple LTE modeling with a
radiative-equilibrium model (as remains common practice in stellar
spectroscopy) would declare this the highest-reaching one, sampling
the coolest upper-atmosphere layers.  However, at the same time it is
the weakest in terms of equivalent width.  The tabulation of
\citetads{1966sst..book.....M} 
gives only 564\,m\AA\ for the Na line, compared to 1259\,m\AA\ for the
Mg line and 3670\,m\AA\ for the Ca line.  The spread of damping wings
also suggests that the Mg line is stronger than the Na line, even
though its core reaches less deep.

\Halpha\ (equivalent width 4020\,m\AA) is shown for comparison.  Its
core is much wider than the others and even shallower than the Mg
line.  The large core width comes from large thermal
broadening at the small hydrogen mass, making it a good
temperature diagnostic
(\citeads{2009A&A...503..577C}). 

The cores of the Na and Mg lines appear similar apart from the larger
Na line depth and the deeper onset of damping wings in the Mg line.
The Ca core is markedly wider and displays marked asymmetry, whereas
the other lines show good symmetry.  That this is not due to blending
is shown by Figure~\ref{fig:neckelcaii} which combines the five
strongest solar \CaII\ lines in the same comparison-enabling format.
All five display similar red-blue core asymmetry.  For \HK\ it appears
as the \KtwoV/\KtwoR\ asymmetry between the slight emission features
that set \HK\ apart from all other Fraunhofer lines in the visible as
being the only ones not showing a simple absorption bell shape.  The
\HK\ asymmetries in spatially-averaged profiles are mostly due to
\HtwoV\ and \KtwoV\ internetwork grains (\cf\ review by
\citeads{1991SoPh..134...15R}). 
These have been explained by Carlsson \& Stein
(\citeyrads{1994chdy.conf...47C}, 
\citeyrads{1997ApJ...481..500C}) 
as acoustic shock interference, taking place in quiet-Sun
clapotispheric gas below the fibrilar \Halpha\ chromosphere (\cf\
\citeads{1995ESASP.376a.151R}). 
Similar but field-guided shocks in near-network dynamic fibrils
(\citeads{2006ApJ...647L..73H}; 
\citeads{2007ApJ...655..624D}) 
may contribute similar asymmetry to the atlas profiles.
Figure~\ref{fig:neckelcaii} suggests that the \CaII\ infrared triplet
lines gain their asymmetry likewise.  We return to this issue in
Section~\ref{sec:shocks} and conclude from this section that:
\begin{itemize}

\item the Mg and Na lines have similar cores but different wings;

\item the imaging asymmetries in Figure~\ref{fig:august} are not
  accompanied by obvious Mg or Na mean-profile asymmetries;

\item only the Ca line displays marked mean-profile asymmetry, which
  it shares with the other strong \CaII\ lines and therefore
  may be due to clapotispheric shocks above the photosphere.

\end{itemize}

\def\figlabel{fig:falcprofiles}    
\begin{figure}       
  \centering
  \includegraphics[width=70mm]{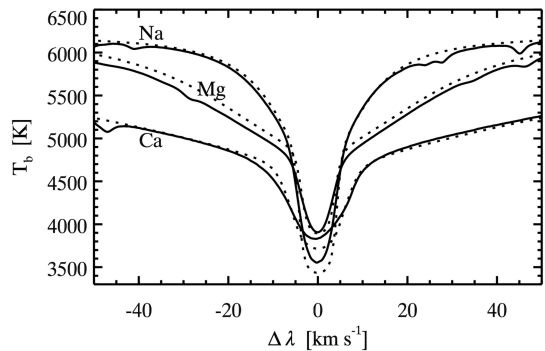} 
  \caption[]{\label{\figlabel} %
      {\em Solid\/}: the three lines in the FTS atlas, as in
    Figure~\ref{fig:neckelred}.  {\em Dotted\/}: emergent line
    profiles for the FALC model in the same units. }
\end{figure}

\def\figlabel{fig:falcformation}
\begin{figure}       
  \centering
  \includegraphics[width=80mm]{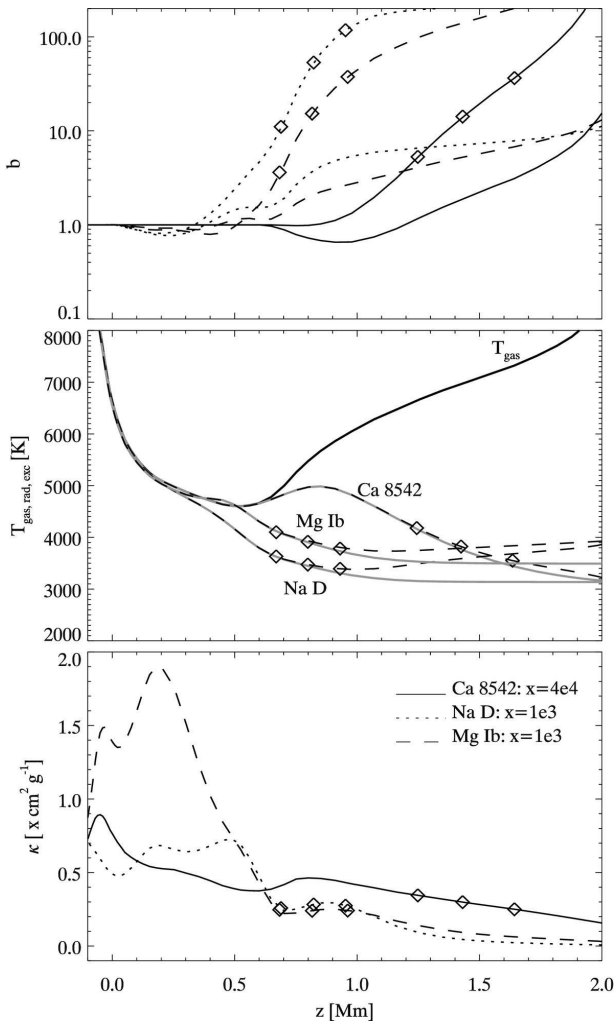} 
  \caption[]{\label{\figlabel} %
    Formation parameters of the three lines in the standard 1D model
    atmosphere of \citetads{1993ApJ...406..319F} 
    against geometrical height.  The symbols mark the heights where
    the total optical depth at line center reaches 0.3, 1 and 3,
    respectively.  {\em Top panel\/}: NLTE population departure
    coefficients $b$ for the Mg line ({\em dashed\/}), Na line ({\em
      dotted\/}) and Ca line ({\em solid\/}).  The optical depth marks
    are superimposed on the lower-level curves.  %
    {\em Middle panel\/}: total (opacity-weighted line plus continuum)
    source function in the form of a formal temperature $T_{\rm exc}$
    ({\em dashed\/}) and mean radiation field in the form of a formal
    temperature $T_{\rm rad}$ ({\em solid\/}) for each of the three
    lines.  The gas temperature $T_{\rm gas}$ is added for
    comparison. %
    {\em Bottom panel\/}: scaled height variation of the total (line
    plus continuum) opacity per gram $\kappa$ at the center of each
    line.  The curve coding is the same as in the top panel.  Note the
    compressed scale for the Ca line; its peak at $z \is 0$ is 24
    times the corresponding peak for the Mg line. }
\end{figure}

\subsection{Formation in the FALC atmosphere} \label{sec:FALC}
After LTE Eddington-Barbier interpretation, the next stage in solar
spectrum modeling is to relax the assumption of LTE but maintain the
assumption of hydrostatic equilibrium in a plane-parallel
simplification.  Figure~\ref{fig:falcprofiles} shows results from such
standard NLTE modeling using the standard FALC model of
\citetads{1993ApJ...406..319F} 
together with our model atoms in the MULTI spectral synthesis code.
The FTS atlas profiles of the three lines are added for comparison.
The pertinent photospheric and low-chromosphere parts of FALC are
identical to those of the quiet-Sun models of
\citetads{1985cdm..proc...67A} 
and \citetads{1986ApJ...306..284M} 
and represent an update of the canonical VAL3C model of
\citetads{1981ApJS...45..635V}. 
They were defined to fit solar continua, not these particular lines,
but Figure~\ref{fig:falcprofiles} shows that they do a rather good job
for these as well.  Our NLTE FALC modeling therefore serves well to
discuss basic formation aspects of our lines.

Figure~\ref{fig:falcformation} diagnoses the FALC formation of the
three lines by plotting various line formation parameters against
height.  The first panel plots population departure coefficients $b
\equiv n/n_\mathrm{LTE}$ where $n$ is the computed population and
$n_\mathrm{LTE}$ the population that would follow from the Saha and
Boltzmann equations from the abundance, gas temperature and electron
density.  They demonstrate that LTE holds to within 20\% in the first
few hundred kilometers of the photosphere.  Above the FALC temperature
minimum (shown by the $T_{\rm gas}$ curve in the center panel) the
lower-level curves show steep increases with height that mark NLTE
underionization in pertinent ionization edges.  In these the
ionization is dominated by photospheric radiation from below without
following the FALC chromospheric temperature rise.  \CaII\
photo-ionization produces Ca lower-level overpopulation only at larger
height.

Deeper down radiative overionization causes small lower-level \MgI\
and \NaI\ underpopulation dips in the upper photosphere\footnote{More
  recently, \citetads{2009ApJ...707..482F} 
  claimed that the much deeper upper-photosphere $b$ dips for \SiII\
  and the other electron-donor species in their newer models are due
  to irradiation from above, not from below, but this is a
  misinterpretation due to using the Menzel definition for departure
  coefficients, see page~37 of
  \citet{2003rtsa.book.....R}. 
  We use the definition of
  \citetads{1972SoPh...23..265W} 
  which normalizes $b$ by the element abundance instead of the
  next-ion density.}.  They are less deep than most atomic species had
in the VAL3C model (\eg\ \FeI\ in
\citeads{1988ASSL..138..185R}) 
for two reasons.  First, the FALC upper photosphere has a less steep
temperature decline, and second, for these levels such dips are filled
in by photon suction, as shown by
\citetads{1992A&A...265..237B} 
for \NaI.  This process of line-driven recombination also operates in
\MgI\ and causes the \MgI\ 12-micron emission features
(\citeads{1992A&A...253..567C}).  
It produces the early onsets of the overpopulation rises in
Figure~\ref{fig:falcformation}, earliest for the Na line.

For each line the upper-level departure coefficient (lower curve of
each pair in the top panel) drops significantly below the lower-level
one through resonance scattering.  The dominance of scattering is
obvious from the center panel in which the three source functions are
close to the angle-averaged radiation field (in $T_{\rm exc}$ and
$T_{\rm rad}$ incarnations to remove Planck function sensitivity
differences).  Since $S^l \approx (b_u/b_l) \, B$ (in the Wien limit),
the divergences between the lower- and upper-level curves in the top
panel define the divergences between the $T_{\rm gas}$ and $T_{\rm
  exc}$ curves in the center panel.
\citetads{1992A&A...265..268U} 
have shown that the textbook case of two-level scattering is an
excellent approximation for the Na line which is a resonance line.
The others are not and may have larger multi-level coupling
to the temperature, for the Ca line via \HK\ (see
\citeads{1980ApJ...241..448O}; 
\citeads{1989A&A...213..360U}). 

The Na source function drops deepest because it uncouples earliest
from the Planck function and feels the FALC chromospheric
temperature rise least.  The Ca source function retains a hump from
this rise and would display small inner-wing emission features in FALC
modeling if the latter wouldn't apply the considerable VAL3C
microturbulence. 

The NLTE source functions and $\tau$ scaling explain the computed
FALC line profiles in Figure~\ref{fig:falcprofiles}.  Application of
the Eddington-Barbier approximation to the $\tau \is 1$ marks in the
center panel recovers the computed line depths.

The bottom panel of Figure~\ref{fig:falcformation} shows the buildup
of the $\tau$ scales, which differs between the three lines (note the
scale factors needed to overplot these curves within one panel).
Plotting the opacity per gram takes out the density dependence.  The
curve patterns are the product of the Saha and Boltzmann population
formulas, the shape of the line extinction profile, and the
lower-level population departures $b_l$ in the top panel.

Let us first discuss the $\kappa$ curve for the Na line (dotted).  The
lower-level population/density ratio $n_l/\rho$ of the Na line in LTE
is set by the ionization equilibrium since this line is from the
ground state and has no Boltzmann excitation sensitivity to
temperature.  This ratio (not shown) has an initial rise due to
decreasing ionization in the initially steeply declining temperature,
followed by a long decline due to larger ionization at lower electron
density.  The photospheric wiggle pattern in the actual Na-line
$\kappa$ curve is added by the $b_l$ variations in the top
panel and by changes in the line profile shape.  The steep $b_l$ rise
halts the ionization decay near $z \is 0.7$~Mm.

The Mg line has three times larger opacity in the lower photosphere
but a much larger drop across the upper photosphere due to the
sensitivity of its Boltzmann excitation to the declining temperature.
The Na and Mg FALC opacities are nearly identical above $h \is
0.5$~Mm.

The Ca line is from the dominant calcium ionization stage; ionization
of \CaII\ to \CaIII\ affects $\kappa_{\rm Ca}$ from $z \is 0.9$~Mm.  In the
photosphere this line has no ionization sensitivity, only excitation
sensitivity to the declining temperature.  This sensitivity is smaller
than for the Mg line (1.7~eV versus 2.7~eV excitation energy).

The wiggles in the deepest layers are set by the steep outward
decrease of the electron/gas density ratio due to hydrogen
recombination.  It affects the ionization and also the balance between
Stark and Van der Waals broadening in setting the broad wings of the
Ca and Mg lines.  The damping constant of the Mg line is about 1.7
times that of the Na line across the FALC photosphere and much larger
in the Stark domain below $h\is0$~Mm.  

We conclude from our FALC modeling:
\begin{itemize}

\item the Mg line has much larger opacity and somewhat larger damping
  than the Na line in the lower photosphere, hence its more prominent
  damping wings in Figure~\ref{fig:neckelred} and its twice-larger
  equivalent width;

\item its Boltzmann sensitivity to the declining temperature across
  the photosphere reduces the Mg line opacity to that of the Na line,
  so that they have nearly identical $\tau \is 1$ heights in FALC;

\item although the FALC chromosphere is not a realistic model of the
  actual chromosphere, this higher-layer opacity equality suggests
  that chromospheric fibrils can have similar opaqueness in the Mg and
  Na lines, as observed in Figure~\ref{fig:august};

\item the various contributions to the $\kappa$ variations around $h
  \is 65$~km in Figure~\ref{fig:falcformation} with their different
  temperature and density sensitivities produce the contrast
  differences along the first and last rows of
  Figure~\ref{fig:simcrispsstsamples}.  For example, the Ca line has
  lower contrast because it is not sensitive to ionization while its
  Boltzmann excitation causes contrast reduction (\cf\ Figure~7 of
  \citeads{2006A&A...449.1209L}). 
  The Mg and Na lines gain contrast from the $\kappa$ sensitivities to
  ionization that produce dips at this height in
  Figure~\ref{fig:falcformation}.

\item the Mg-line opacity has larger temperature sensitivity than the
  Na-line opacity due to its excitation energy.  This may contribute
  to the larger clapotisphere transparency for Mg that is seen in the
  center row of Figure~\ref{fig:simsamples};

\item the Ca line has much larger opacity across all of FALC, which
  explains that it also maps chromospheric fibrils better than the
  other two lines.  In the upper photosphere the Ca wings sample
  identical scenes as the other two lines but further out from line
  center, as seen in
  Figures~\ref{fig:simsamples}--\ref{fig:simcrispsstsamples}.

\item difference in resonance scattering explains that the Na
  line-center image in Figure~\ref{fig:obssamples} appears the
  fuzziest, the Ca line-center image the sharpest.  Similar sharpness
  difference is seen between the line-center Dopplergrams in
  Figure~\ref{fig:doppsamples}, also for the simulation.

\end{itemize}

\def\figlabel{fig:napixdemo}    
\begin{figure}       
  \centering
  \includegraphics[width=70mm]{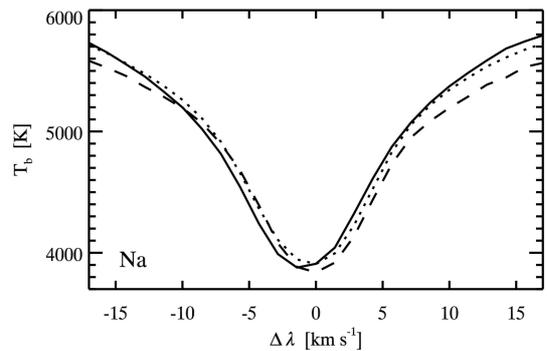} 
  \caption[]{\label{\figlabel} %
    Representative observed Na-line profiles to illustrate the cause
    of imaging asymmetries in granulation and reversed
    granulation. {\em Solid\/}: pixel at the center of a granule.
    {\em Dashed\/}: pixel in an intergranular lane. {\em
      Dotted\/}: spatial average.}
\end{figure}

\subsection{Granulation and reversed granulation} \label{sec:revgran}
Comparison of the second and next-to-last rows in
Figures~\ref{fig:obssamples}--\ref{fig:simcrispsstsamples} confirms
the imaging asymmetries in Figure~\ref{fig:august}.
Figure~\ref{fig:neckelred} shows that it does not have to do with
mean-profile asymmetry.  Figure~\ref{fig:falcformation} demonstrates
that the Na line and Mg line have near-equal FALC core formation.
Therefore, the reason for the imaging asymmetry must be dynamics that
is not included in the FALC modeling.  The Dopplergrams in
Figure~\ref{fig:doppsamples} demonstrate that the reversed-granulation
scenes in Figures~\ref{fig:obssamples}--\ref{fig:simcrispsstsamples}
are fully set by granular dynamics.  The latter causes the imaging
asymmetries.

Figure~\ref{fig:napixdemo} is a simple demonstration for the Na line
following the cartoon for \BaII\,4554\,\AA\ in Figure~6 of
\citetads{2001A&A...378..251S} 
with actual data.  (Similar cartoons were already shown in Figure~4 of
\citeads{1964ApNr....9...33E}.) 
The lane pixel has a redshifted profile that is darker than
the mean profile, especially in the outer wings.  The granule pixel
has a profile close to the spatial mean but slightly blueshifted and
slightly brighter in the outer wings.  A filtergram near $\Delta
\lambda = -14$~\kms\ as the first Na panel in
Figures~\ref{fig:obssamples}--\ref{fig:simcrispsstsamples} shows
``flattened'' granulation with smaller contrast than between
nonshifted profiles.  The reverse holds for the opposite sampling at
$\Delta \lambda = +14$~\kms.  This explains that in
Figures~\ref{fig:obssamples}--\ref{fig:simcrispsstsamples} the top Na
panel shows lower granulation contrast than the bottom Na panel.  The
corresponding Dopplershift panels in Figure~\ref{fig:doppsamples} show
the granulation even clearer, both for the observation and the
simulation.

The same comparison for the Ca line shows no such difference, the
corresponding observed Dopplergram only noise.  The reason is simply
that the comparable scene (or height) sampling takes place much
further out in the extended damping wings of the Ca line, where these
have a shallow slope so that profile shift has less effect.  This is also
the case for the Mg line which differs from the Na line in possessing
extended wings (Figure~\ref{fig:neckelred}).

Let us now make such comparisons closer to the core.  The blue wings
of the lane and granule profiles in Figure~\ref{fig:napixdemo} cross,
giving granulation contrast reversal which does not occur on the red
side.  These two pixels show reversed granulation with reversed
imaging asymmetry.  The effect is again larger for steeper line
flanks.  The corresponding sampling temperature of about 5000~K lies
above the onset of the Mg and Ca wings in
Figure~\ref{fig:falcprofiles}.  Therefore, the striking imaging
asymmetry between the wings of the Na line in Figure~\ref{fig:august}
is also simply due to their steepness (Figure~\ref{fig:neckelred}).
The same asymmetry should appear in the inner core of other narrow but
somewhat weaker lines; indeed it is very pronounced for \FeI\,7090.4\,\AA\
in Figure~5 of
\citetads{2006A&A...450..365J}. 

The diffuse-gray lane patterns in the middle-row Dopplergrams in
Figure~\ref{fig:doppsamples} correspond well to the vertical velocity
pattern at $z\is0.3$~Mm in the last row of Figure~5 of
\citetads{2010ApJ...709.1362L} 
and qualitatively to the similar pattern at $\tau=0.1$ in Figure~3 of
\citetads{2007A&A...461.1163C}. 
Such simulations also show slight temperature contrast reversal
between granules and lanes at such heights, in our case illustrated
for $z\is0.3$~Mm in the second row of Figure~5 of
\citetads{2010ApJ...709.1362L}. 
It was already explained by
\citetads{1984ssdp.conf..181N} 
on pages 15--18 in his informative analysis of his pioneering
granulation simulation as due to radiative reheating along bend-over
flow trajectories, as those in Figure~6 of
\citetads{2007A&A...461.1163C}. 
The Dopplershift-brightening pattern combination produces slender
near-lane brightness features as seen in the second row of
Figure~\ref{fig:obssamples}.  They correlate only roughly with
underlying intergranular lanes, best at a delay of 1--2 minutes.  In
addition, at larger height gravity wave interference takes over from
direct granular overshoot and reduces the correspondence (\cf\
\citeads{2004A&A...416..333R}; 
\citeads{2005A&A...431..687L}; 
\citeads{2006A&A...450..365J}; 
\citeads{2008ApJ...681L.125S}). 

The conclusion from this section is that the imaging asymmetries
between the Na wings and between the Mg and Na lines in
Figure~\ref{fig:august} arise from the correlated brightness and
Dopplershift patterns in the solar granulation and its overshoot, to
which the Na line is more sensitive because it does not have strong
damping wings.

\def\figlabel{fig:Hinode} 
\begin{figure*}      
  \sidecaption
  \includegraphics[width=12cm]{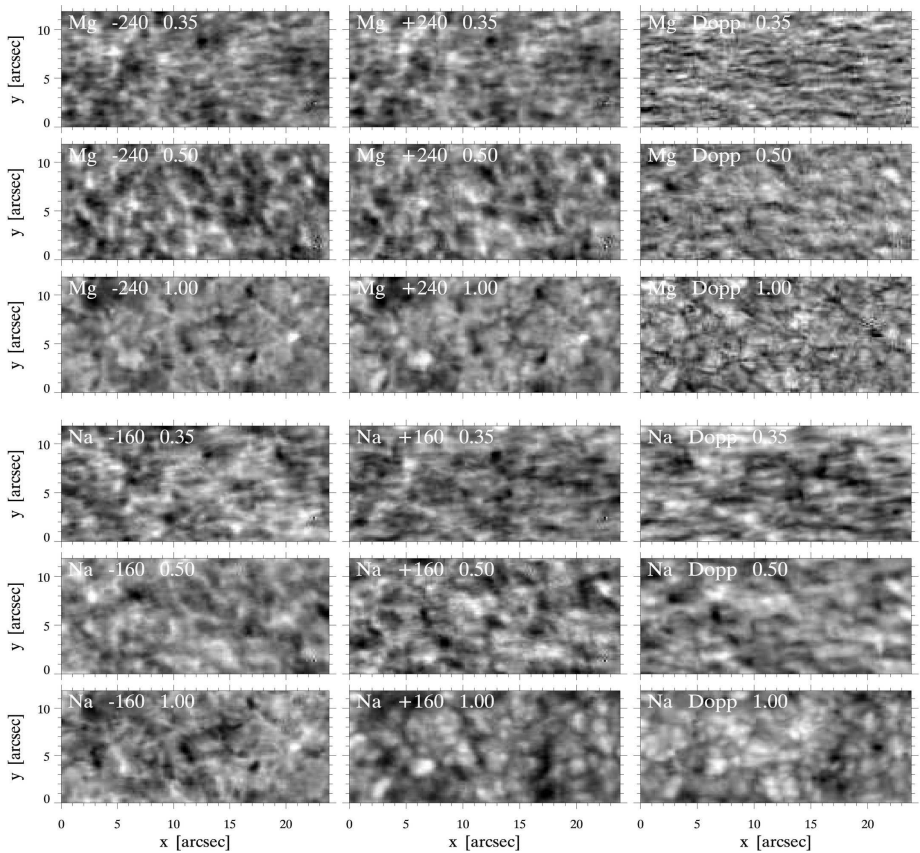}
  \caption[]{\label{\figlabel} %
    Comparable intensity images from Hinode in the Mg line ({\em upper
      half}) and Na line ({\em lower half}).  Each panel is
    independently grayscaled for optimum contrast.  The subfield has
    the same size as in
    Figures~\ref{fig:obssamples}--\ref{fig:doppsamples}.  Along rows
    the images cover the same scene, with about 32~s delay
    between the two lines.  The viewing angle differs between rows,
    with $\mu \is 0.35$, 0.50 and 1.00 as specified.  The limb is to
    the top.  The first two columns are images in the blue and red
    wing at $\Delta \lambda \is \pm 240$\,m\AA\ for the Mg line,
    $\Delta \lambda \is \pm 160$\,m\AA\ for the Mg line.  The third
    column has the corresponding $(I_{\rm red} - I_{\rm blue})/(I_{\rm
      red} + I_{\rm blue})$ Dopplergrams. \\[1ex] }
\end{figure*}

\subsection{Line core scenes}  \label{sec:cores}

All three lines are scattering lines, the Na line the worst.  Their
core intensities are set by radiation escape in deeper layers
than the last scatterings.  The latter set the Dopplershift encoding
(assuming complete redistribution which is in order even for the Na
Line, see \citeads{1992A&A...265..268U}). 
This sampling difference is most clearly demonstrated by the simulated
Mg core, which shows regular granulation much clearer in intensity
than in its Dopplergram.  The difference also explains that the large
image variations between the three central rows of
Figure~\ref{fig:obssamples}--\ref{fig:simcrispsstsamples} are largely
produced by the Dopplershifts in the bottom rows of
Figure~\ref{fig:doppsamples}.  Therefore, taking Dopplergrams
represents a much better use of these lines than taking intensity
filtergrams.  

The rough correspondence of the bright and dark Mg and Na Dopplergram
patches in the third row of Figure~\ref{fig:doppsamples} with the
areas of brighter and less bright granules in the first row suggests
that these are $p$-mode oscillation patches that result from global
interference patterning, with larger $D$ in the blueshifted phase.
The lambdameter phase difference measurements for \NaDtwo\ of
\citetads{1996A&A...307..936D} 
indeed show oscillation coherence between the outer wing and core, but
only for the five-minute component (last panel of their Figure 6).
For increasing height the oscillation dominance shifts from
five-minute evanescent global mode interference to more local
three-minute propagating waves, as seen already in the somewhat deeper
layers sampled by ultraviolet continua (\eg\
\citeads{2001A&A...379.1052K}), 
and then to three-minute shocks that are seen in the Ca line (next
section).  The lack of coherence at higher frequency in Figure~6 of
\citetads{1996A&A...307..936D} 
suggests that the Mg and Na core Dopplergrams already have some
sensitivity to these shocks.

The simulation does not contain global $p$-mode oscillation
interference patterning but has its own box modes and generates its
own propagating waves.  The fine structure across the last row of
Figure~\ref{fig:doppsamples} shows that all three lines Doppler-sense
the very inhomogeneous shock-ridden clapotisphere that results, even
though it appears too transparent in the simulated Mg line
(Section~\ref{sec:discussion}).  We refer to Leenaarts \etal\
(\citeyrads{2009ApJ...694L.128L}, 
\citeyrads{2010ApJ...709.1362L}) 
for pertinent displays.  In particular, Figure~9 of the latter paper
contains informative formation breakdown diagrams for the Na and Ca
lines.  Comparing these to Figure~\ref{fig:falcformation} demonstrates
enormous line formation variations with respect to FALC modeling.  

The conclusion of this section is that the line cores primarily sample
wave patterns that likely include shock interference.

\subsection{Magnetic concentrations and shocks}   \label{sec:shocks}
There are only a few magnetic concentrations in the observed subfield,
more in the simulation.  In
Figures~\ref{fig:obssamples}--\ref{fig:simcrispsstsamples} they stand
out as bright features in the outer line wings, often brighter than
the brightest granules because lack of collisional damping in the
partially evacuated concentrations reduces the wing opacity (\cf\
Leenaarts \etal\
\citeyrads{2006A&A...452L..15L}, 
\citeyrads{2006A&A...449.1209L}). 

Nearer the line centers they no longer stand out as bright points but
rather as profile asymmetry features, producing small, roundish, dark
features in the Mg and Na Dopplergrams in
Figure~\ref{fig:doppsamples}.  Figure~\ref{fig:bscatter} uses the
$V/R$ asymmetry ratio
following the example for \CaII\ \HtwoV\ grains of
\citetads{1983ApJ...272..355C}. 
The formation diagrams of such grains in
\citetads{1997ApJ...481..500C} 
show that \HtwoV\ grain asymmetry results from line-of-sight
integration through cool postshock downflows around $h \is\ 1.5$~Mm
and grain-producing upcoming hot fresh shocks around $h \is 1$~Mm.
The same pattern occurs for rarefied magnetic concentrations at
heights of only a few hundred kilometers, where the outside scene is
still reversed granulation.  The ubiquity of photospheric shocks in
and near isolated magnetic concentrations was first suggested by
\citetads{1998ApJ...495..468S}. 
\rev{Their signature in the Na core was also noted by
  \citetads{2010ApJ...719L.134J}.} 

A detailed demonstration is given by the Na-line formation diagrams in
the first quartet of Figure~7 of
\citetads{2010ApJ...709.1362L}, 
which analyze the magnetic concentration at $(x,y)\is(8.2,7.2)$ in the
simulation snapshot.  It has a shock at $z \approx 0.3$~Mm that
produces an emission peak just blue of Na line center, in the same
manner as a \HtwoV\ grain but much deeper.  The postshock downdrafts
from shocks that previously passed, now at $z \approx 0.5$ and 1.0~Mm,
produce a red-shifted line core which is very dark because the
strongly scattering Na source function decouples from the Planck
function already at $h \approx 0.2$~Mm and shows no sensitivity to the
actual high-temperature plateau over $h \approx 0.3 - 0.7$~Mm.  Such
raised temperature plateaus in magnetic concentrations are attributed
to Joule heating by
\citetads{2010MmSAI..81..582C}. 
They are similar to the FALC chromospheric temperature plateau but
occur at photospheric heights.

Note that when such a spectral emission peak blueshifts into an
inner-wing sampling passband, \rev{the Dopplergram sign 
of $D \equiv (T_\rmb^{\rm red} - T_\rmb^{\rm
      blue})/ (T_\rmb^{\rm red} + T_\rmb^{\rm blue})$}
becomes the same as
for a redshifted absorption core. \KtwoV-like grains can therefore
appear black in our $D$ definition.  Similarly, the asymmetry ratio
$V/R$ of Figure~\ref{fig:bscatter} may exceed unity both due to
fresh-shock bright updraft and to dark post-shock downfall, \ie\ for
both phases of shock modulation and especially when these combine
along the line of sight.

The Ca image just blue of line center in the third row of
Figure~\ref{fig:obssamples} does show non-magnetic bright grains that
represent the Ca-line analog to higher-formed internetwork \KtwoV\
grains.  Their visibility is to be expected from the similar core
asymmetries in all five \CaII\ lines in Figure~\ref{fig:neckelcaii}.
They are indeed black in the corresponding Dopplergram in
Figure~\ref{fig:doppsamples}.  The center-row Ca diagram in
Figure~\ref{fig:bscatter} suggests larger spatial filling factor for
these shocks.  The simulation diagram even has a secondary contour
peak for them.

Outside magnetic concentrations the Na and Mg lines are too weak to
produce internetwork grains.  Nevertheless, the Na-core $V/R$
distribution tail in the central diagram of Figure~\ref{fig:bscatter}
suggests that the Na line senses internetwork shocks similarly to the
Ca line.

The conclusion of this section is that Na and Mg core Dopplergrams or
asymmetry images turn out to be quasi-magnetograms by mapping
photospheric shocks in or near strong-field concentrations.  Outside
these, only the Ca line shows mean-profile asymmetry from
clapotispheric internetwork shocks.

\section{Center-to-limb variation}  \label{sec:Hinode}

Figure~\ref{fig:Hinode} presents a selection from the Hinode
observations, added to show how limbward viewing affects the three
lines.  The $\Delta \lambda$ values are selected for scene similarity
at $\mu\is1$.  These images (bottom rows of each half) have
lower quality than the SST/CRISP images but again show similar imaging
asymmetry as in Figure~\ref{fig:august}: similar reversed granulation
in the blue Mg and Na wings and the red Mg wing, granular-like
patterning instead in the red Na wing.  The $\mu \is 1$ Dopplergrams
also show granulation, with larger contrast in the Na line.

At $\mu \is 0.5$ the pattern repeats but with less discrepancy for the
Na red wing, less difference between the Dopplergrams, and general
foreshortening.  The decrease in imaging asymmetry corresponds to
smaller Dopplershifts from the vertical motions.

Closer to the limb a different pattern of horizontal stripings
appears, similarly in all $\mu \is 0.35$ panels and clearest in the Mg
Dopplergram.  It looks rather like the supergranulation in the
discovery plate in Figure~6 of
\citetads{1962ApJ...135..474L} 
but of course the scale is much smaller: these are horizontal outflows
in granulation cells rather than in supergranulation cells.  At this
formation height they probably correspond to the bend-over flow lines
of \citetads{2007A&A...461.1163C}. 
In the Na Dopplergram some dark patches are visible which are likely
clapotispheric clouds that are not visible in the Mg line due to lack of
Boltzmann excitation.

\def\figlabel{fig:doppcenter} 
\begin{figure*}
  \center
  \includegraphics[width=18cm]{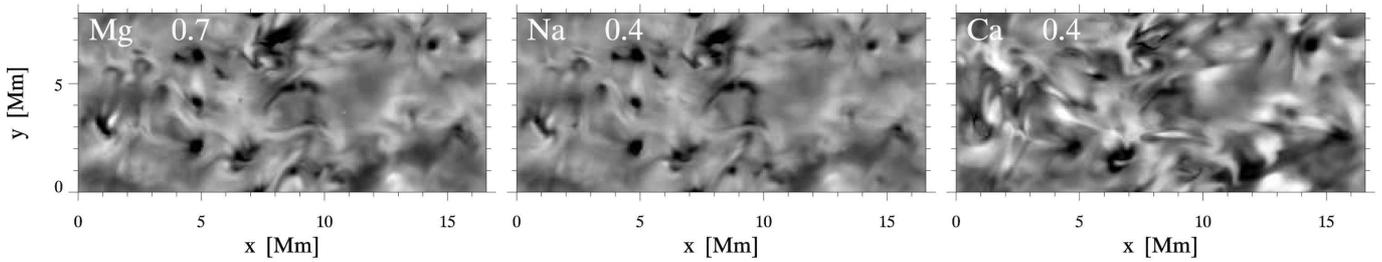}
  \caption[]{\label{\figlabel} %
    Dopplergrams from the simulation as in the
    bottom row of Figure~\ref{fig:doppsamples} but closer to line
    center and with only CRISP spectral smearing, no spatial PSF
    smearing. \\[1ex] }
\end{figure*}

\section{Discussion}                                \label{sec:discussion}

In Figures~\ref{fig:obssamples}--\ref{fig:doppsamples} we chose to
jump from deep outer-wing formation ($h\approx150$~km) in the second
rows to line-core formation in the next rows to avoid figure clutter
and to skip observation-simulation sampling differences set by the too
narrow simulated Ca and Mg cores. In the observations the intervening
scenes, best seen per Dopplergram, are simply the combination of the
(reversed) granulation and overlying larger-scale oscillation patches.
The cloud-like appearance of the latter suggests bimodal formation,
with the deeper layers well described by Eddington-Barbier sampling
near $\tau \is 1$, the upper layers better described by
Schuster-Schwarschild clouds with Dopplershift as the most important
cloud parameter.  The inner-core panels in Figure~6 of
\citetads{1996A&A...307..936D} 
confirm the notion of cloud-like common Dopplershift at all
frequencies.  In the simulation the visibility of a gray
low-Dopplershift intermediate background scene that appears especially
in the outer Ca core is likely due to incomplete masking by
clapotispheric clouds that are too transparent.

Cauzzi \etal\
(\citeyrads{2009A&A...503..577C}, 
\citeyrads{2008A&A...480..515C}) 
suggested such bimodal formation for chromospheric fibrils in the Ca
line.  These indeed blanket the inner Ca core even more when there is
more activity than in the very quiet area analysed here.  In the
Balmer lines they even tend to obscure the reversed granulation
(\citeads{2006A&A...449.1209L}). 
Note that
\citetads{1972SoPh...27..299E} 
already reported not finding negative ``oddities'' for \Hbeta\ (but
note also that
\citetads{1972SoPh...27..303T} 
as referee blamed fibril source functions, not opacities).

The simulation snapshot and line synthesis do a good job in
reproducing and so helping to explain the imaging asymmetry between
the Na line and the other two lines, but they do not reproduce the
observations in the line cores very well.  These are not the main
topic here so we discuss these discrepancies only briefly, again
referring to Leenaarts \etal\
(\citeyrads{2009ApJ...694L.128L}, 
\citeyrads{2010ApJ...709.1362L}) 
for more detail.  The major ones are the larger visibility of
granulation at the center of the Mg line, the narrower core width of
the Ca line in the simulation, and the differences in core asymmetry
sampling in Figure~\ref{fig:bscatter}.

The clapotispheric cool-cloud transparency in the simulated Mg line is
likely an artifact due to underestimation of the \MgI\ density, for
the same reason as the underestimation of the \NaI\ density in these
cool clouds discussed in
\citetads{2010ApJ...709.1362L}: 
since in the simulation the ionization equilibria are assumed to be in
LTE, the electron-donor elements (Si, Fe, Al, Mg) are less ionized
than they would be if the actual domination by photoionization was
properly accounted for.  The resulting underestimation of the electron
density is 1--2 orders of magnitude, higher up even 3--4 if one
compares with time-dependent hydrogen ionization balancing in cool
post-shock gas (\citeads{2007A&A...473..625L}). 
In contrast, the synthesis codes implement photoionization properly,
so these yield higher ionization rates, but the recombination lacks
electrons so that the balance shifts to the ion state and the line
opacity is severely underestimated.  Thus, the cool clouds are too
transparent in the simulated Na and Mg lines, more so in the latter
due to its 2.7~eV excitation energy.  Possibly the Ca line is
similarly affected higher up in cool clouds by similarly overestimated
ionization from \CaII\ to \CaIII.  Such transparencies may contribute
to the differences in Figure~\ref{fig:bscatter}.

The too narrow core width of the Ca line was noted and discussed by
\citetads{2009ApJ...694L.128L}. 
The apparent need for ``microturbulent broadening'' in higher layers
suggests a lack of numerical resolution, just as the solar granulation
needed microturbulent fudging until simulation resolution did away
with that (Nordlund 
\citeyrads{1980LNP...114..213N}, 
\citeyrads{1984ssdp.conf..181N}). 
Indeed, a test with the higher-resolution simulation of
\citetads{2010MmSAI..81..582C} 
shows larger broadening.  The assumption of instantaneous LTE
ionization balancing not only causes underestimation of the electron
densities in cool clouds and corresponding underestimation of the
opacities of our three lines, but also of the clapotispheric dynamics
(\citeads{2007A&A...473..625L}). 

These deficiencies suggest that the actual clapotispheric
contributions to these line cores may be even more wildly structured
than Figures~\ref{fig:simcrispsstsamples}--\ref{fig:doppsamples}
suggest already.  Figure~\ref{fig:doppcenter} displays innermost-core
Dopplergrams from the simulation after only spectral smearing with the
CRISP transmission functions.  The spectral smearing affects the core
depth considerably (Figure~\ref{fig:sampleprofiles}) but the
Dopplergram scene keeps its structure.  The scene is a violent one
with sharp bright fronts, dark cool clouds, and black magnetic
concentrations.  Comparison of Figure~\ref{fig:doppcenter} to the Na
line-center Dopplershift panel in Figure~7 of
\citetads{2010ApJ...709.1362L} 
shows that the CRISP-smeared Dopplergram gives a good rendering of the
actual Dopplershifts, with more emphasis on magnetic concentrations.
Comparison of Figure~\ref{fig:doppcenter} to the center rows of
Figure~\ref{fig:simsamples} shows that the simulated line cores are so
much dominated by Doppler modulation that inner-core Doppler
measurement is mandatory in using these lines as upper-atmosphere
diagnostics.  Comparison of Figure~\ref{fig:doppcenter} to the
line-core rows of Figure~\ref{fig:doppsamples} suggests that higher
spatial resolution than in our observations is highly desirable.

A corollary is that searches for global $g$-modes and so-called
``chromosphere seismology'' using full-disk resonance-cell sampling of
the Na line as with GOLF-NG (\eg\
\citeads{2006AdSpR..38.1812T}; 
\citeads{2009ASPC..416..341S}) 
will suffer noise from reversed granulation just as classical
helioseismology suffers noise from granulation.  Magnetic
concentrations contribute much noise (literally) by their shocks.
GOLF-NG's 15 passbands, spread over $\Delta \lambda \is \pm 9$~\kms,
have $\mbox{FWHM} \is 30$\,m\AA, half CRISP's passband in this line
and clearly narrow enough.  Interpretation of such multi-passband
oscillation sampling that relies on one-dimensional
height-of-formation interpretation, for example to diagnose upward
propagation from inward phase difference, is likely to fail since the
clapotispheric signals are better described as clouds of varying
opacity at varying height of which the varying Dopplershifts act as
shutters obscuring the inner line wings.

\section{Conclusion}                               \label{sec:conclusion}
There is a striking imaging asymmetry between \NaDone\ and \Mgbtwo\ in
Figure~\ref{fig:august}: the Mg line shows reversed granulation in
both wings, but the Na line shows similar reversed granulation only in
its blue wing, normal granulation in its red wing.  This is due to
different sensitivity to the correlated Dopplershift and brightness
pattern of the reversed granulation in the mid-photosphere.  The Mg
line has larger opacity and damping in deeper layers, hence wider
damping wings, hence less steep profile flanks, hence smaller Doppler
sensitivity.  Both lines sample the same layers in both their wings;
only the appearance of the granulation and the reversed granulation
differ between these, not their physical nature or sampling height.

The cores of the Mg and Na lines are formed remarkably similarly.
Dopplergrams in these represent their primary diagnostic value and can
also serve to detect and study magnetic concentrations marked by
photospheric shocks.

Center-to-limb observations with at least the quality of the
disk-center SST/CRISP data used here will be well-suited to constrain
the reversed-granulation domain further and to enlarge the testing
ground for numerical simulations.  

When the lack of line-core opacity and broadening in the simulation is
resolved, the obvious next step is to compare time sequences of such
imaging spectroscopy and simulations for these lines in Fourier
analyses as those of \eg\
\citetads{1989A&A...224..245F} 
and
\citetads{1996A&A...307..936D}. 

The small structural scales that are predicted by the simulations to
dominate at clapotispheric heights define the need for high spatial
resolution, considerably higher even than reached here, as a principal
quest in probing this domain of the solar atmosphere.

\begin{acknowledgements}
  We thank T.D.~Tarbell and P.~S\"utterlin for the discussions that
  inspired this paper and T.~Leifsen, H.~Skogsrud and A.~Ortiz for
  obtaining the CRISP scan in Figure~\ref{fig:august} that initiated
  these discussions.  The SST is operated by The Institute for Solar
  Physics of the Royal Swedish Academy of Sciences at the Spanish
  Observatorio del Roque de los Muchachos of the Instituto de
  Astrof\'{\i}sica de Canarias.  The Japanese Hinode mission was
  developed and launched by ISAS/JAXA, with NAOJ as domestic partner
  and NASA and STFC as international partners, and is operated by
  these agencies in co-operation with ESA and NSC.  RJR's travel to
  the SST and to Oslo was funded by the EC through the SOLAIRE Network
  (MTRN-CT-2006-035484).  Discussions at NAOJ (Mitaka, Japan) while
  RJR was a visitor there also contributed to this paper, and he
  completed it as a visitor at LMSAL (Palo Alto, USA).  As always, our
  work was much facilitated by NASA's ADS.
\end{acknowledgements}


\begin{thebibliography}{81}
\expandafter\ifx\csname natexlab\endcsname\relax\def\natexlab#1{#1}\fi

\bibitem[{{Altrock} \& {Musman}(1976)}]{1976ApJ...203..533A}
{Altrock}, R.~C. \& {Musman}, S. 1976, \apj, 203, 533

\bibitem[{{Avrett}(1985)}]{1985cdm..proc...67A}
{Avrett}, E.~H. 1985, in Chromospheric Diagnostics and Modelling, ed.
  {B.~W.~Lites}, 67

\bibitem[{{Bard} \& {Carlsson}(2008)}]{2008ApJ...682.1376B}
{Bard}, S. \& {Carlsson}, M. 2008, \apj, 682, 1376

\bibitem[{{Botnen}(1997)}]{1997MsT..........2B}
{Botnen}, A. 1997, Master's thesis, 
  Inst.~Theor.~Astrophys.~Oslo

\bibitem[{{Bruls} {et~al.}(1992){Bruls}, {Rutten}, \&
  {Shchukina}}]{1992A&A...265..237B}
{Bruls}, J.~H.~M.~J., {Rutten}, R.~J., \& {Shchukina}, N.~G. 1992, \aap, 265,
  237

\bibitem[{{Canfield} \& {Mehltretter}(1973)}]{1973SoPh...33...33C}
{Canfield}, R.~C. \& {Mehltretter}, J.~P. 1973, \solphys, 33, 33

\bibitem[{{Carlsson}(1986)}]{1986UppOR..33.....C}
{Carlsson}, M. 1986, Uppsala Astron.\ Obs.\ Reports, 33

\bibitem[{{Carlsson} {et~al.}(2010){Carlsson}, {Hansteen}, \&
  {Gudiksen}}]{2010MmSAI..81..582C}
{Carlsson}, M., {Hansteen}, V.~H., \& {Gudiksen}, B.~V. 2010, \memsai, 81, 582

\bibitem[{{Carlsson} {et~al.}(1992){Carlsson}, {Rutten}, \&
  {Shchukina}}]{1992A&A...253..567C}
{Carlsson}, M., {Rutten}, R.~J., \& {Shchukina}, N.~G. 1992, \aap, 253, 567

\bibitem[{{Carlsson} \& {Stein}(1994)}]{1994chdy.conf...47C}
{Carlsson}, M. \& {Stein}, R.~F. 1994, in Chromospheric Dynamics, ed.
  {M.~Carlsson}, 47

\bibitem[{{Carlsson} \& {Stein}(1997)}]{1997ApJ...481..500C}
{Carlsson}, M. \& {Stein}, R.~F. 1997, \apj, 481, 500

\bibitem[{{Cauzzi} {et~al.}(2009){Cauzzi}, {Reardon}, {Rutten}, {Tritschler},
  \& {Uitenbroek}}]{2009A&A...503..577C}
{Cauzzi}, G., {Reardon}, K., {Rutten}, R.~J., {Tritschler}, A., \&
  {Uitenbroek}, H. 2009, \aap, 503, 577

\bibitem[{{Cauzzi} {et~al.}(2008){Cauzzi}, {Reardon}, {Uitenbroek},
  {Cavallini}, {Falchi}, {Falciani}, {Janssen}, {Rimmele}, {Vecchio}, \&
  {W{\"o}ger}}]{2008A&A...480..515C}
{Cauzzi}, G., {Reardon}, K.~P., {Uitenbroek}, H., {et~al.} 2008, \aap, 480, 515

\bibitem[{{Cheung} {et~al.}(2007){Cheung}, {Sch{\"u}ssler}, \&
  {Moreno-Insertis}}]{2007A&A...461.1163C}
{Cheung}, M.~C.~M., {Sch{\"u}ssler}, M., \& {Moreno-Insertis}, F. 2007, \aap,
  461, 1163

\bibitem[{{Collados} \& {V{\'{a}}zquez}(1987)}]{1987A&A...180..223C}
{Collados}, M. \& {V{\'{a}}zquez}, M. 1987, \aap, 180, 223

\bibitem[{{Collet} {et~al.}(2005){Collet}, {Asplund}, \&
  {Th{\'e}venin}}]{2005A&A...442..643C}
{Collet}, R., {Asplund}, M., \& {Th{\'e}venin}, F. 2005, \aap, 442, 643

\bibitem[{{Cram} \& {Dam{\'{e}}}(1983)}]{1983ApJ...272..355C}
{Cram}, L.~E. \& {Dam{\'{e}}}, L. 1983, \apj, 272, 355

\bibitem[{{De Pontieu} {et~al.}(2007){De Pontieu}, {Hansteen}, {Rouppe van der
  Voort}, {van Noort}, \& {Carlsson}}]{2007ApJ...655..624D}
{De Pontieu}, B., {Hansteen}, V.~H., {Rouppe van der Voort}, L., {van Noort},
  M., \& {Carlsson}, M. 2007, \apj, 655, 624

\bibitem[{{Deubner} {et~al.}(1996){Deubner}, {Waldschik}, \&
  {Steffens}}]{1996A&A...307..936D}
{Deubner}, F., {Waldschik}, T., \& {Steffens}, S. 1996, \aap, 307, 936

\bibitem[{{Deubner} \& {Mattig}(1975)}]{1975A&A....45..167D}
{Deubner}, F.~L. \& {Mattig}, W. 1975, \aap, 45, 167

\bibitem[{{Evans}(1964)}]{1964ApNr....9...33E}
{Evans}, J.~W. 1964, Astrophysica Norvegica, 9, 33

\bibitem[{{Evans} \& {Catalano}(1972)}]{1972SoPh...27..299E}
{Evans}, J.~W. \& {Catalano}, C.~P. 1972, \solphys, 27, 299

\bibitem[{{Fleck} \& {Deubner}(1989)}]{1989A&A...224..245F}
{Fleck}, B. \& {Deubner}, F. 1989, \aap, 224, 245

\bibitem[{{Fleck} {et~al.}(2010){Fleck}, {Straus}, {Carlsson}, {Jefferies},
  {Severino}, \& {Tarbell}}]{2010MmSAI..81..777F}
{Fleck}, B., {Straus}, T., {Carlsson}, M., {et~al.} 2010, \memsai, 81, 777

\bibitem[{{Fontenla} {et~al.}(1993){Fontenla}, {Avrett}, \&
  {Loeser}}]{1993ApJ...406..319F}
{Fontenla}, J.~M., {Avrett}, E.~H., \& {Loeser}, R. 1993, \apj, 406, 319

\bibitem[{{Fontenla} {et~al.}(2009){Fontenla}, {Curdt}, {Haberreiter},
  {Harder}, \& {Tian}}]{2009ApJ...707..482F}
{Fontenla}, J.~M., {Curdt}, W., {Haberreiter}, M., {Harder}, J., \& {Tian}, H.
  2009, \apj, 707, 482

\bibitem[{{Hansteen} {et~al.}(2007){Hansteen}, {Carlsson}, \&
    {Gudiksen}}]{2007ASPC..368..107H} {Hansteen}, V.~H., {Carlsson},
  M., \& {Gudiksen}, B. 2007, in \aspcs\ 368, The Physics of
  Chromospheric Plasmas, ed. {P.~Heinzel, I.~Dorotovi{\v c}, \&
    R.~J.~Rutten}, 107

\bibitem[{{Hansteen} {et~al.}(2006){Hansteen}, {De Pontieu}, {Rouppe van der
  Voort}, {van Noort}, \& {Carlsson}}]{2006ApJ...647L..73H}
{Hansteen}, V.~H., {De Pontieu}, B., {Rouppe van der Voort}, L., {van Noort},
  M., \& {Carlsson}, M. 2006, \apjl, 647, L73

\bibitem[{{Janssen} \& {Cauzzi}(2006)}]{2006A&A...450..365J}
{Janssen}, K. \& {Cauzzi}, G. 2006, \aap, 450, 365

\bibitem[{{Jess} {et~al.}(2010){Jess}, {Mathioudakis}, {Christian}, {Crockett},
  \& {Keenan}}]{2010ApJ...719L.134J}
{Jess}, D.~B., {Mathioudakis}, M., {Christian}, D.~J., {Crockett}, P.~J., \&
  {Keenan}, F.~P. 2010, \apjl, 719, L134

\bibitem[{{Kosugi} {et~al.}(2007){Kosugi}, {Matsuzaki}, {Sakao}, {Shimizu},
  {Sone}, {Tachikawa}, {Hashimoto}, {Minesugi}, {Ohnishi}, {Yamada}, {Tsuneta},
  {Hara}, {Ichimoto}, {Suematsu}, {Shimojo}, {Watanabe}, {Shimada}, {Davis},
  {Hill}, {Owens}, {Title}, {Culhane}, {Harra}, {Doschek}, \&
  {Golub}}]{2007SoPh..243....3K}
{Kosugi}, T., {Matsuzaki}, K., {Sakao}, T., {et~al.} 2007, \solphys, 243, 3

\bibitem[{{Krijger} {et~al.}(2001){Krijger}, {Rutten}, {Lites}, {Straus},
  {Shine}, \& {Tarbell}}]{2001A&A...379.1052K}
{Krijger}, J.~M., {Rutten}, R.~J., {Lites}, B.~W., {et~al.} 2001, \aap, 379,
  1052

\bibitem[{{Leenaarts} \& {Carlsson}(2009)}]{2009ASPC..415...87L}
  {Leenaarts}, J. \& {Carlsson}, M. 2009, in \aspcs\ 415, The Second
  Hinode Science Meeting: Beyond Discovery-Toward Understanding,
  ed. {B.~Lites, M.~Cheung, T.~Magara, J.~Mariska, \& K.~Reeves}, 87

\bibitem[{{Leenaarts} {et~al.}(2009){Leenaarts}, {Carlsson}, {Hansteen}, \&
  {Rouppe van der Voort}}]{2009ApJ...694L.128L}
{Leenaarts}, J., {Carlsson}, M., {Hansteen}, V., \& {Rouppe van der Voort}, L.
  2009, \apjl, 694, L128

\bibitem[{{Leenaarts} {et~al.}(2007){Leenaarts}, {Carlsson}, {Hansteen}, \&
  {Rutten}}]{2007A&A...473..625L}
{Leenaarts}, J., {Carlsson}, M., {Hansteen}, V., \& {Rutten}, R.~J. 2007, \aap,
  473, 625

\bibitem[{{Leenaarts} {et~al.}(2006{\natexlab{a}}){Leenaarts}, {Rutten},
  {Carlsson}, \& {Uitenbroek}}]{2006A&A...452L..15L}
{Leenaarts}, J., {Rutten}, R.~J., {Carlsson}, M., \& {Uitenbroek}, H.
  2006{\natexlab{a}}, \aap, 452, L15

\bibitem[{{Leenaarts} {et~al.}(2010){Leenaarts}, {Rutten}, {Reardon},
  {Carlsson}, \& {Hansteen}}]{2010ApJ...709.1362L}
{Leenaarts}, J., {Rutten}, R.~J., {Reardon}, K., {Carlsson}, M., \& {Hansteen},
  V. 2010, \apj, 709, 1362

\bibitem[{{Leenaarts} {et~al.}(2006{\natexlab{b}}){Leenaarts}, {Rutten},
  {S{\"u}tterlin}, {Carlsson}, \& {Uitenbroek}}]{2006A&A...449.1209L}
{Leenaarts}, J., {Rutten}, R.~J., {S{\"u}tterlin}, P., {Carlsson}, M., \&
  {Uitenbroek}, H. 2006{\natexlab{b}}, \aap, 449, 1209

\bibitem[{{Leenaarts} \& {Wedemeyer-B{\"o}hm}(2005)}]{2005A&A...431..687L}
{Leenaarts}, J. \& {Wedemeyer-B{\"o}hm}, S. 2005, \aap, 431, 687

\bibitem[{{Leighton} {et~al.}(1962){Leighton}, {Noyes}, \&
  {Simon}}]{1962ApJ...135..474L}
{Leighton}, R.~B., {Noyes}, R.~W., \& {Simon}, G.~W. 1962, \apj, 135, 474

\bibitem[{{Levy}(1971)}]{1971A&A....14...15L}
{Levy}, M. 1971, \aap, 14, 15

\bibitem[{Lockyer(1868)}]{Lockyer1868}
Lockyer, N. 1868, Procs.\ Royal Soc.\ London, 17, 131

\bibitem[{{Maltby} {et~al.}(1986){Maltby}, {Avrett}, {Carlsson},
  {Kjeldseth-Moe}, {Kurucz}, \& {Loeser}}]{1986ApJ...306..284M}
{Maltby}, P., {Avrett}, E.~H., {Carlsson}, M., {et~al.} 1986, \apj, 306, 284

\bibitem[{{Moore} {et~al.}(1966){Moore}, {Minnaert}, \&
  {Houtgast}}]{1966sst..book.....M}
{Moore}, C.~E., {Minnaert}, M.~G.~J., \& {Houtgast}, J. 1966, {The solar
  spectrum 2935~\AA\ to 8770~\AA}

\bibitem[{Neckel(1999)}]{Neckel1999}
Neckel, H. 1999, Sol.\ Phys., 184, 421

\bibitem[{{Neckel} \& {Labs}(1984)}]{1984SoPh...90..205N}
{Neckel}, H. \& {Labs}, D. 1984, \solphys, 90, 205

\bibitem[{{Nordlund}(1980)}]{1980LNP...114..213N} {Nordlund}, A. 1980,
  IAU Colloq.\ 51: Stellar Turbulence, ed. {D.~F.~Gray \&
    J.~L.~Linsky}, Lecture Notes in Physics, Springer, 114, 213

\bibitem[{{Nordlund}(1984{\natexlab{a}})}]{1984ssdp.conf..174N}
{Nordlund}, A. 1984{\natexlab{a}}, in Small-Scale Dynamical Processes in Quiet
  Stellar Atmospheres, ed. {S.~L.~Keil}, 174

\bibitem[{{Nordlund}(1984{\natexlab{b}})}]{1984ssdp.conf..181N}
{Nordlund}, A. 1984{\natexlab{b}}, in Small-Scale Dynamical Processes in Quiet
  Stellar Atmospheres, ed. {S.~L.~Keil}, 181

\bibitem[{{Owocki} \& {Auer}(1980)}]{1980ApJ...241..448O}
{Owocki}, S.~P. \& {Auer}, L.~H. 1980, \apj, 241, 448

\bibitem[{{Roddier}(1981)}]{1981PrOpt..19..281R}
{Roddier}, F. 1981, Prog.~Optics, 19, 281

\bibitem[{{Rouppe van der Voort} {et~al.}(2009){Rouppe van der Voort},
  {Leenaarts}, {De Pontieu}, {Carlsson}, \& {Vissers}}]{2009ApJ...705..272R}
{Rouppe van der Voort}, L., {Leenaarts}, J., {De Pontieu}, B., {Carlsson}, M.,
  \& {Vissers}, G. 2009, \apj, 705, 272

\bibitem[{{Rutten}(1988)}]{1988ASSL..138..185R} {Rutten}, R.~J. 1988,
  in IAU Colloq.\ 94: Physics of Formation of FE II Lines Outside LTE,
  ed. {R.~Viotti, A.~Vittone, \& M.~Friedjung}, ASSL, Springer, 138,
  185

\bibitem[{{Rutten}(1995)}]{1995ESASP.376a.151R} {Rutten}, R.~J. 1995,
  in ESA Special Pub.\ 376, Helioseismology, 151

\bibitem[{{Rutten}(2003)}]{2003rtsa.book.....R} {Rutten}, R.~J. 2003,
  {Radiative Transfer in Stellar Atmospheres}, Lecture notes Utrecht
  University

\bibitem[{Rutten(2010)}]{Rutten2010b} Rutten, R.~J. 2010, in Recent
  Advances in Spectroscopy: Astrophysical, Theoretical and
  Experimental Perspectives, ed. R.~K. Chaudhuri, M.~V.  Mekkaden,
  A.~V. Raveendran, \& A.~S. Narayanan, Astrophys.\ Space Science
  Procs., Springer, 163, preprint 2009arXiv0912.2206L

\bibitem[{{Rutten} {et~al.}(2004){Rutten}, {De Wijn}, \&
    {S{\"u}tterlin}}]{2004A&A...416..333R} {Rutten}, R.~J., {De Wijn},
  A.~G., \& {S{\"u}tterlin}, P. 2004, \aap, 416, 333

\bibitem[{{Rutten} \& {Uitenbroek}(1991)}]{1991SoPh..134...15R}
{Rutten}, R.~J. \& {Uitenbroek}, H. 1991, \solphys, 134, 15

\bibitem[{{Salabert} {et~al.}(2009){Salabert}, {Turck-Chi{\`e}ze},
    {Barri{\`e}re}, {Carton}, {Daniel-Thomas}, {Delbart},
    {Garc{\'{\i}}a}, {Granelli}, {Jim{\'e}nez-Reyes},
    {Lahonde-Hamdoun}, {Loiseau}, {Mathur}, {Nunio}, {Pall{\'e}},
    {Piret}, {Robillot}, \& {Simoniello}}]{2009ASPC..416..341S}
  {Salabert}, D., {Turck-Chi{\`e}ze}, S., {Barri{\`e}re}, J.~C.,
  {et~al.} 2009, in \aspcs\ 416, Solar-Stellar Dynamos as Revealed by
  Helio- and Asteroseismology: GONG 2008/SOHO 21, ed. {M.~Dikpati,
    T.~Arentoft, I.~Gonz{\'a}lez Hern{\'a}ndez, C.~Lindsey, \&
    F.~Hill}, 341

\bibitem[{{Scharmer} {et~al.}(2003{\natexlab{a}}){Scharmer}, {Bjelksjo},
  {Korhonen}, {Lindberg}, \& {Petterson}}]{2003SPIE.4853..341S}
{Scharmer}, G.~B., {Bjelksjo}, K., {Korhonen}, T.~K., {Lindberg}, B., \&
  {Petterson}, B. 2003{\natexlab{a}}, 
  SPIE Conf.\ Series 4853, 341

\bibitem[{{Scharmer} {et~al.}(2003{\natexlab{b}}){Scharmer},
    {Dettori}, {L{\"o}fdahl}, \& {Shand}}]{2003SPIE.4853..370S}
  {Scharmer}, G.~B., {Dettori}, P.~M., {L{\"o}fdahl}, M.~G., \&
  {Shand}, M.  2003{\natexlab{b}}, SPIE Conf.\ Series 4853, 370

\bibitem[{{Scharmer} {et~al.}(2008){Scharmer}, {Narayan}, {Hillberg}, {de la
  Cruz Rodrigu{\'{e}}z}, {L{\"o}fdahl}, {Kiselman}, {S{\"u}tterlin}, {van
  Noort}, \& {Lagg}}]{2008ApJ...689L..69S}
{Scharmer}, G.~B., {Narayan}, G., {Hillberg}, T., {et~al.} 2008, \apjl, 689,
  L69

\bibitem[{{Sch{\"u}ssler} {et~al.}(2003){Sch{\"u}ssler}, {Shelyag},
  {Berdyugina}, {V{\"o}gler}, \& {Solanki}}]{2003ApJ...597L.173S}
{Sch{\"u}ssler}, M., {Shelyag}, S., {Berdyugina}, S., {V{\"o}gler}, A., \&
  {Solanki}, S.~K. 2003, \apjl, 597, L173

\bibitem[{{Shelyag} {et~al.}(2004){Shelyag}, {Sch{\"u}ssler}, {Solanki},
  {Berdyugina}, \& {V{\"o}gler}}]{2004A&A...427..335S}
{Shelyag}, S., {Sch{\"u}ssler}, M., {Solanki}, S.~K., {Berdyugina}, S.~V., \&
  {V{\"o}gler}, A. 2004, \aap, 427, 335

\bibitem[{{Steiner} {et~al.}(1998){Steiner}, {Grossmann-Doerth},
  {Kn{\"{o}}lker}, \& {Sch{\"{u}}ssler}}]{1998ApJ...495..468S}
{Steiner}, O., {Grossmann-Doerth}, U., {Kn{\"{o}}lker}, M., \&
  {Sch{\"{u}}ssler}, M. 1998, \apj, 495, 468

\bibitem[{{Straus} {et~al.}(2008){Straus}, {Fleck}, {Jefferies}, {Cauzzi},
  {McIntosh}, {Reardon}, {Severino}, \& {Steffen}}]{2008ApJ...681L.125S}
{Straus}, T., {Fleck}, B., {Jefferies}, S.~M., {et~al.} 2008, \apjl, 681, L125

\bibitem[{{Suematsu} {et~al.}(2008){Suematsu}, {Tsuneta}, {Ichimoto},
  {Shimizu}, {Otsubo}, {Katsukawa}, {Nakagiri}, {Noguchi}, {Tamura}, {Kato},
  {Hara}, {Kubo}, {Mikami}, {Saito}, {Matsushita}, {Kawaguchi}, {Nakaoji},
  {Nagae}, {Shimada}, {Takeyama}, \& {Yamamuro}}]{2008SoPh..249..197S}
{Suematsu}, Y., {Tsuneta}, S., {Ichimoto}, K., {et~al.} 2008, \solphys, 249,
  197

\bibitem[{{Suemoto} {et~al.}(1987){Suemoto}, {Hiei}, \&
  {Nakagomi}}]{1987SoPh..112...59S}
{Suemoto}, Z., {Hiei}, E., \& {Nakagomi}, Y. 1987, \solphys, 112, 59

\bibitem[{{Suemoto} {et~al.}(1990){Suemoto}, {Hiei}, \&
  {Nakagomi}}]{1990SoPh..127...11S}
{Suemoto}, Z., {Hiei}, E., \& {Nakagomi}, Y. 1990, \solphys, 127, 11

\bibitem[{{S{\"u}tterlin} {et~al.}(2001){S{\"u}tterlin}, {Rutten}, \&
  {Skomorovsky}}]{2001A&A...378..251S}
{S{\"u}tterlin}, P., {Rutten}, R.~J., \& {Skomorovsky}, V.~I. 2001, \aap, 378,
  251

\bibitem[{{Thomas}(1972)}]{1972SoPh...27..303T}
{Thomas}, R.~N. 1972, \solphys, 27, 303

\bibitem[{{Tsuneta} {et~al.}(2008){Tsuneta}, {Ichimoto}, {Katsukawa}, {Nagata},
  {Otsubo}, {Shimizu}, {Suematsu}, {Nakagiri}, {Noguchi}, {Tarbell}, {Title},
  {Shine}, {Rosenberg}, {Hoffmann}, {Jurcevich}, {Kushner}, {Levay}, {Lites},
  {Elmore}, {Matsushita}, {Kawaguchi}, {Saito}, {Mikami}, {Hill}, \&
  {Owens}}]{2008SoPh..249..167T}
{Tsuneta}, S., {Ichimoto}, K., {Katsukawa}, Y., {et~al.} 2008, \solphys, 249,
  167

\bibitem[{{Turck-Chi{\`e}ze} {et~al.}(2006){Turck-Chi{\`e}ze}, {Carton},
  {Ballot}, {Barri{\`e}re}, {Daniel-Thomas}, {Delbart}, {Desforges},
  {Garc{\'{\i}}a}, {Granelli}, {Mathur}, {Nunio}, {Piret}, {Pall{\'e}},
  {Jim{\'e}nez}, {Jim{\'e}nez-Reyes}, {Robillot}, {Fossat}, {Eff-Darwich}, \&
  {Gelly}}]{2006AdSpR..38.1812T}
{Turck-Chi{\`e}ze}, S., {Carton}, P., {Ballot}, J., {et~al.} 2006, Adv.\ 
  Space Res.\ 38, 1812

\bibitem[{{Uitenbroek}(1989)}]{1989A&A...213..360U}
{Uitenbroek}, H. 1989, \aap, 213, 360

\bibitem[{{Uitenbroek} \& {Bruls}(1992)}]{1992A&A...265..268U}
{Uitenbroek}, H. \& {Bruls}, J.~H.~M.~J. 1992, \aap, 265, 268

\bibitem[{{van Noort} {et~al.}(2005){van Noort}, {Rouppe van der Voort}, \&
  {L{\"o}fdahl}}]{2005SoPh..228..191V}
{van Noort}, M., {Rouppe van der Voort}, L., \& {L{\"o}fdahl}, M.~G. 2005,
  \solphys, 228, 191

\bibitem[{{van Noort} \& {Rouppe van der Voort}(2008)}]{2008A&A...489..429V}
{van Noort}, M.~J. \& {Rouppe van der Voort}, L.~H.~M. 2008, \aap, 489, 429

\bibitem[{{Vernazza} {et~al.}(1981){Vernazza}, {Avrett}, \&
  {Loeser}}]{1981ApJS...45..635V}
{Vernazza}, J.~E., {Avrett}, E.~H., \& {Loeser}, R. 1981, \apjs, 45, 635

\bibitem[{{Wedemeyer-B{\"o}hm}(2008)}]{2008A&A...487..399W}
{Wedemeyer-B{\"o}hm}, S. 2008, \aap, 487, 399

\bibitem[{{Wedemeyer-B{\"o}hm} \& {Rouppe van der
  Voort}(2009)}]{2009A&A...503..225W}
{Wedemeyer-B{\"o}hm}, S. \& {Rouppe van der Voort}, L. 2009, \aap, 503, 225

\bibitem[{{Wijbenga} \& {Zwaan}(1972)}]{1972SoPh...23..265W}
{Wijbenga}, J.~W. \& {Zwaan}, C. 1972, \solphys, 23, 265

\end{thebibliography}

\end{document}